\documentclass[sigconf]{acmart}

\usepackage{balance}       
\usepackage{graphics}      
\usepackage{color}
\usepackage{booktabs}
\usepackage{textcomp}
\usepackage{subcaption}
\usepackage{makecell}
\usepackage{multicol}
\usepackage{multirow}
\usepackage{array}
\usepackage{enumitem}
\usepackage{amsmath}
\usepackage{stfloats}
\usepackage{graphicx}
\usepackage{amsthm}
\usepackage{listings}
\usepackage{caption} 
\usepackage{xspace}
\usepackage[linesnumbered]{algorithm2e}

\newcommand*{\eg}{\textit{e.g.},\xspace}
\newcommand*{\ie}{\textit{i.e.},\xspace}
\newcommand*{\vs}{\textit{v.s.}\xspace}
\newcommand*{\etc}{\textit{etc.}\xspace}

\newcommand*{\etal}{\textit{et~al.}\xspace}

\newenvironment{s_itemize}{
\begin{itemize}[leftmargin=*]
  \setlength{\itemsep}{3pt}
  \setlength{\parskip}{0pt}
  \setlength{\parsep}{0pt}
}{\end{itemize}}

\newenvironment{s_enumerate}{
\begin{enumerate}[leftmargin=*]
  \setlength{\itemsep}{2pt}
  \setlength{\parskip}{0pt}
  \setlength{\parsep}{0pt}
}{\end{enumerate}}

\definecolor{DarkGreen}{HTML}{5DAC81}
\newcommand\review[1]{\textcolor{black}{#1}}

\AtBeginDocument{%
  \providecommand\BibTeX{{%
    \normalfont B\kern-0.5em{\scshape i\kern-0.25em b}\kern-0.8em\TeX}}}

\copyrightyear{2022}
\acmYear{2022}
\setcopyright{acmcopyright}\acmConference[CHI '22]{CHI Conference on Human
Factors in Computing Systems}{April 29-May 5, 2022}{New Orleans, LA, USA}
\acmBooktitle{CHI Conference on Human Factors in Computing Systems (CHI '22),
April 29-May 5, 2022, New Orleans, LA, USA}
\acmPrice{15.00}
\acmDOI{10.1145/3491102.3501904}
\acmISBN{978-1-4503-9157-3/22/04}

\begin{document}

\title{Enabling Hand Gesture Customization on Wrist-Worn Devices}
\subtitle{\xspace}

\author{Xuhai Xu}
\affiliation{%
  \institution{Apple Inc. and UW} \city{} \state{} \country{}
}
\email{xuhaixu@uw.edu}

\author{Jun Gong}
\affiliation{%
  \institution{Apple Inc.} \city{} \state{} \country{}
}
\email{jun_gong@apple.com}

\author{Carolina Brum}
\affiliation{%
  \institution{Apple Inc.} \city{} \state{} \country{}
}
\email{brum@apple.com}

\author{Lilian Liang}
\affiliation{%
  \institution{Apple Inc.} \city{} \state{} \country{}
}
\email{lilian_liang@apple.com}

\author{Bongsoo Suh}
\affiliation{%
  \institution{Apple Inc.} \city{} \state{} \country{}
}
\email{bsuh@apple.com}

\author{Shivam Kumar Gupta}
\affiliation{%
  \institution{Apple Inc.} \city{} \state{} \country{}
}
\email{sgupta25@apple.com}

\author{Yash Agarwal}
\affiliation{%
  \institution{Apple Inc.} \city{} \state{} \country{}
}
\email{yash_s_agarwal@apple.com}

\author{Laurence Lindsey}
\affiliation{%
  \institution{Apple Inc.} \city{} \state{} \country{}
}
\email{laurence_lindsey@apple.com}

\author{Runchang Kang}
\affiliation{%
  \institution{Apple Inc.} \city{} \state{} \country{}
}
\email{runchang_kang@apple.com}

\author{Behrooz Shahsavari}
\affiliation{%
  \institution{Apple Inc.} \city{} \state{} \country{}
}
\email{bshahsavari@apple.com}

\author{Tu Nguyen}
\affiliation{%
  \institution{Apple Inc.} \city{} \state{} \country{}
}
\email{tu@apple.com}

\author{Heriberto Nieto}
\affiliation{%
  \institution{Apple Inc.} \city{} \state{} \country{}
}
\email{hnieto@apple.com}

\author{Scott E. Hudson}
\affiliation{%
  \institution{Apple Inc. and CMU} \city{} \state{} \country{}
}
\email{scott.hudson@cs.cmu.edu}

\author{Charlie Maalouf}
\affiliation{%
  \institution{Apple Inc.} \city{} \state{} \country{}
}
\email{cmaalouf@apple.com}

\author{Jax Seyed Mousavi}
\affiliation{%
  \institution{Apple Inc.} \city{} \state{} \country{}
}
\email{hseyedmousavi@apple.com}

\author{Gierad Laput}
\affiliation{%
   \institution{Apple Inc.} \city{} \state{} \country{}
}
\email{gierad@apple.com}


\renewcommand{\shortauthors}{Xu \textit{et al.}}
\renewcommand{\shorttitle}{Gesture Customization}

\begin{abstract}

We present a framework for gesture customization requiring minimal examples from users, all without degrading the performance of existing gesture sets. To achieve this, we first deployed a large-scale study (N=500+) to collect data and train an accelerometer-gyroscope recognition model with a cross-user accuracy of 95.7\% and a false-positive rate of 0.6 per hour when tested on everyday non-gesture data.
Next, we design a few-shot learning framework which derives a lightweight model from our pre-trained model, enabling knowledge transfer without performance degradation. 
We validate our approach through a user study (N=20) examining on-device customization from 12 new gestures, resulting in an average accuracy of 55.3\%, 83.1\%, and 87.2\% on using one, three, or five shots when adding a new gesture, while maintaining the same recognition accuracy and false-positive rate from the pre-existing gesture set.
We further evaluate the usability of our real-time implementation with a user experience study (N=20). Our results highlight the effectiveness, learnability, and usability of our customization framework.
Our approach paves the way for a future where users are no longer bound to pre-existing gestures, freeing them to creatively introduce new gestures tailored to their preferences and abilities.

\end{abstract}



\maketitle

\section{Introduction}
\label{sec:introduction}

\begin{figure*}[t]
    \centering
    \includegraphics[width=1\textwidth]{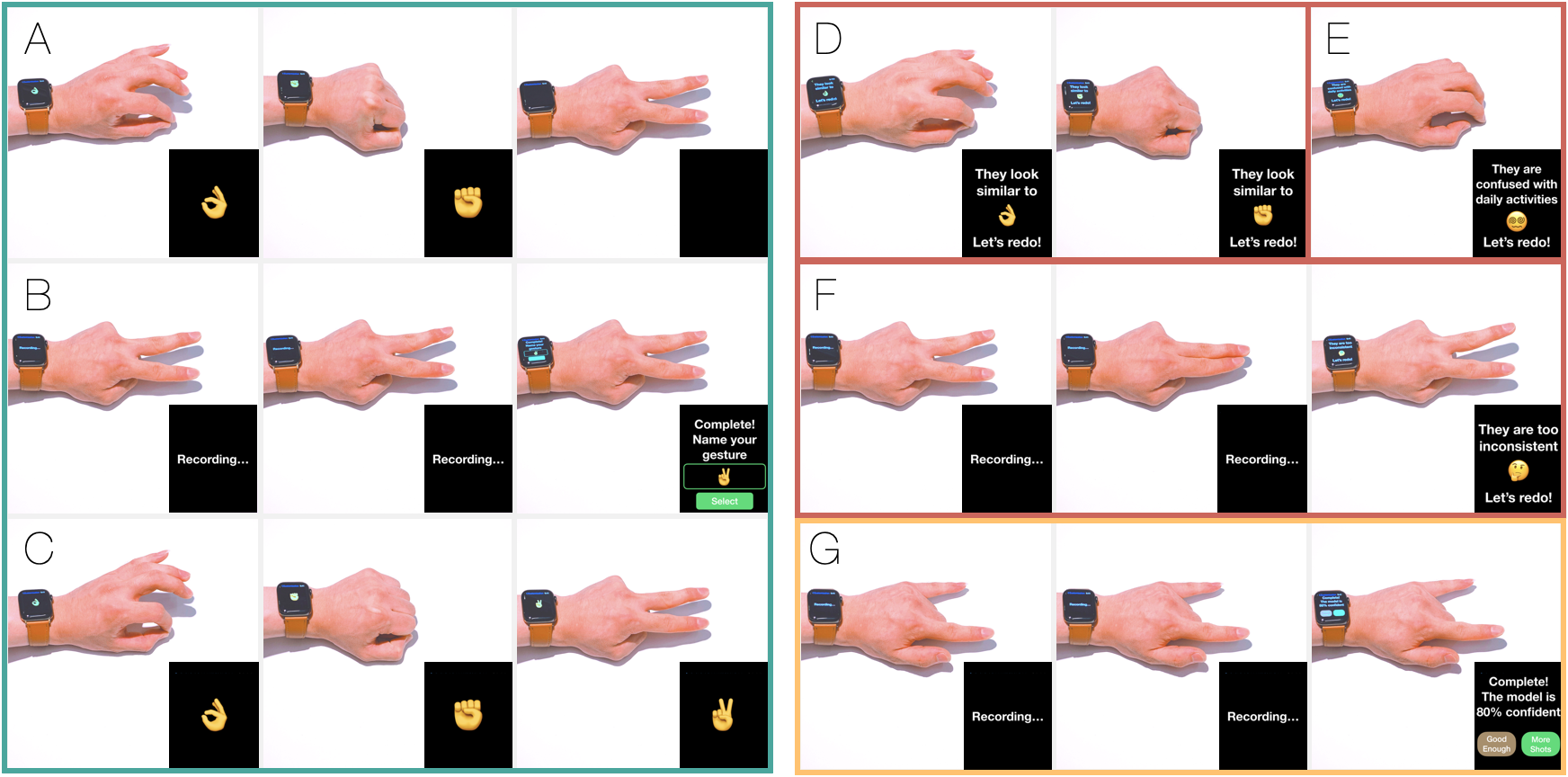}
    \caption{Overview of Our Gesture Customization Framework \& Real-Time System. A model can recognize a pre-existing gesture set and is robust to noise, but it cannot recognize out-of-dictionary gestures (A). A custom gesture can be added merely with three samples (B), without affecting the recognition on pre-existing gestures (C).
    The system can provide real-time feedback when the new gesture is close to existing gestures (D), similar to common daily activities (E), and inconsistent (F). When the model performance is sub-optimal, users can decide whether to provide additional samples (G).}
    \Description{
Seven groups of hand gesture figures, each with a screenshot of the corresponding watch interface.
Group A, B, C are about one procedure.
Group A: pinch and clench gestures are recognized and their results are displayed on the watch, but the peace gesture is not recognized.
Group B: three figures of peace gestures with a recording mode on the watch.
Group C: pinch, clench, and peace can all be recognized now.
Group D: pinch and clench are tried to be added as new gestures but get rejected. The system informs that they are similar to existing gestures.
Group E: a casual pose is tried to be added as a new gesture but gets rejected. The system informs that it is too close to common daily activities.
Group F: three figures of peace gestures with a recording mode on the watch, but the second peace has the two-finger combined, which is quite different from the other two where the two fingers are separate. It is rejected by the system which informs it is too inconsistent.
Group G: three figures of spider-man gestures with a recording mode on the watch. The model has an accuracy of 80\%. The system informs users about this and asks for users to press one of two buttons: "good enough", and "more shots".
    }
    \label{fig:overview_hero}
\end{figure*}

Recent advances in worn sensing technologies have led to the emergence of promising hand gesture recognition systems~\cite{gong_wristwhirl_2016,wu_back-hand-pose_2020,iravantchi_beamband_2019,truong_capband_2018,kim_imu_2019,jung_wearable_2015, escalera_gesture_2017, lien_soli_2016}. Regardless of the sensing modality, typical systems are designed with pre-defined gestures, using data collected from multiple users~\cite{laput_sensing_2019,wen_serendipity_2016,xu_hulamove_2021,zhang_tomo_2015,xu_earbuddy_2020}.
To truly leverage gestural input, devices should allow users to add their own gestures beyond a pre-existing gesture vocabulary~\cite{oh_challenges_2013,lou_personalized_2017,wobbrock_user-defined_2009}. This unlocks several advantages, including better gesture memorability~\cite{nacenta_memorability_2013}, higher interaction efficiency~\cite{ouyang_bootstrapping_2012}, and enhanced accessibility for people with special needs~\cite{anthony_analyzing_2013}. 


However, there are two notable requirements for systems aiming to support gesture customization, especially if the bar is to prevent performance degradation of the original gesture set.
First, such a system should support a rapid and minimal data collection process (\eg around five samples) to limit user burden.
Second, such a system should go beyond model fine-tuning, as this often causes two problems~\cite{tao_few-shot_2020}: one is \textit{forgetting old}, where the model's performance deteriorates drastically on old (\ie pre-defined) classes~\cite{french_catastrophic_1999}; the other is \textit{overfitting new}, where the fine-tuned model is prone to overfitting towards newer classes, therefore degrading its generalizability~\cite{gidaris_dynamic_2018}.

Although several early systems have explored few-shot gesture recognition (first requirement)~\cite{liu_uwave_2009,bigdelou_adaptive_2012}, they mainly work for  simple, salient gestures and rely on highly distinguishable signals ~\cite{liu_uwave_2009,avrahami_guided_2001,anthony_lightweight_2010}. This often leads to poor performance for gestures that are more fine-grained and natural.
\review{Moreover, these systems collapse all gestures (old and new) into a single global gesture set. They did not distinguish the pre-existing and customized gestures. But to make such a system robust in the wild, it is essential to track performance benchmarks between pre-existing \vs customized gesture sets, and doing equally well on both are of paramount importance.
}
To the best of our knowledge, no prior work has investigated the challenges of extending an existing model for new gestures with few shots (second requirement).

In this paper, we propose a robust gesture customization framework that supports a small number of user examples (three to five~\cite{wang_generalizing_2020,rahimian_few-shot_2021}), and offers in-situ feedback while maintaining recognition performance on the original gesture set. Our framework integrates transfer learning, incremental learning, and few-shot learning techniques.
To do this, we first conduct a user study on over 500 users across diverse contexts, using accelerometer-gyroscope data to train a convolution neural network (CNN) to recognize a pre-defined gesture set (four gestures plus a null class, which includes 60 hours of of daily activities such as walking, typing, driving, cooking). 
Next, we train a lightweight model for custom gestures, without adjusting the parameters of the original model. We employ the first half of our pre-trained model as a feature embedding extractor, and create a parallel output after the embedding layer to enable the training of a new model without affecting the pre-trained model.
We then utilize a series of data augmentation, data synthesis, and adversarial training techniques to extract the most utility from few user samples and boost model performance.

To further ensure that a new gesture is reliable \cite{oh_challenges_2013}, we designed our learning and training process around an interactive customization experience. Instead of simply accepting any gesture, our system provides interactive feedback when a new gesture is either 1) \textit{too close} to the existing gesture set, 2) \textit{too inconsistent} relative to provided examples, 3) \textit{too confusing} against unintended interactions such as frequent daily activities, and 4) has \textit{sub-optimal recognition performance}, during which users can decide whether they want to provide additional samples.
Our study results show that this feedback mechanism empowers users to understand and make better use of the recognition system to suit their accuracy and reliability preferences.
Figure~\ref{fig:overview_hero} illustrates the overview of our framework.

To summarize, our paper makes the following core contributions:

\begin{s_itemize}
\item We propose a few-shot gesture customization framework that can support gesture customization with a small number of samples, without degrading the performance of pre-defined gestures.

\item We conducted a large-scale user study (500+ participants) and built a wrist-worn accelerometer-gyroscope gesture recognition model. This model can recognize four gestures with an accuracy of 95.7\% and an F1 score of 95.8\% in a cross-user setup, and a false positive rate of 0.6 times per hour when tested on daily behavior non-gesture data.

\item We extend the pre-trained model by designing architectural, data augmentation, data synthesis, adversarial regularization techniques, and interactive feedback mechanisms to extract the most utility from few user samples. \review{We evaluated our framework on 12 new gestures in addition to the four existing gestures.} The final model achieves an average accuracy of 55.3\%, 83.1\%, and 87.2\%, and an F1 score of 66.0\%, 89.2\%, and 92.1\% on one, three, and five gesture examples.

\item We evaluate the usability of our real-time three-shot gesture customization system through a user study. Our results indicate that our gesture customization system achieves highly favorable usability and learnability effects. We believe our framework enables end-users to easily and creatively introduce new gestures tailored to their preferences and abilities.

\end{s_itemize}
\section{Related Work}
\label{sec:related_work}
In this section, we first provide a general overview of wearable hand gesture recognition techniques across sensing modalities.
Next, we review prior work in the gesture customization. Finally, we review ML-based methods that are relevant to our work.

\subsection{Wearable Hand Gesture Recognition}
\label{sub:related_work:gesture_recognition}
Researchers and practitioners have extensively explored hand gesture recognition. These technologies leverage diverse modalities, including cameras~\cite{kim_digits_2012,wu_back-hand-pose_2020,yeo_opisthenar_2019,hu_fingertrak_2020,xu_hand_2018}, infrared (IR) ranging~\cite{gong_wristwhirl_2016, mcintosh_sensir_2017,georgi_recognizing_2015,fukui_hand_2011, kienzle_lightring_2014}, acoustics~\cite{nandakumar_fingerio_2016,laput_sensing_2019, harrison_skinput_2010,iravantchi_beamband_2019,iravantchi_interferi_2019}, electromyography (EMG)~\cite{saponas_enabling_2009,saponas_demonstrating_2008,caramiaux_understanding_2015}, electrical impedance tomography~\cite{zhang_tomo_2015}, pressure~\cite{dementyev_wristflex_2014,jung_wearable_2015}, radar~\cite{lien_soli_2016}, stretch sensors~\cite{strohmeier_flick_2012}, magnetic sensors~\cite{chen_finexus_2016,chen_utrack_2013,parizi_auraring_2019,yang_magic_2012}, and bio-capacitive effects~\cite{rekimoto_gesturewrist_2001,truong_capband_2018,sato_touche_2012}. Among these techniques, the intertial measurement unit (IMU) is arguably one of the most low-cost and widely available sensors embedded in commodity wearable devices. As a result, the IMU is frequently relied upon for capturing dynamic hand gestures that involve arm or hand motion~\cite{laput_viband_2016,wen_serendipity_2016,xu_finger-writing_2015,akl_novel_2011,kim_imu_2019,escalera_gesture_2017,caramiaux_adaptive_2015}. In this work, we focused on the IMU, specifically the accelerometer and gyroscope, for its ubiquity and potential for generalizability.

Early trajectory-based gesture recognition methods, such as dynamic time warping (DTW)~\cite{liu_uwave_2009} and hidden Markov models (HMM)~\cite{mckenna_comparison_2004}, can recognize gesture trajectories (\eg line, square, circle, star~\cite{mckenna_comparison_2004}) using few samples while achieving high accuracies.
However, these methods do not work well for gestures that are more complex and fine-grained.
More sophisticated techniques have emerged, relying heavily on data-driven approaches. These are typically designed by collecting data from a known set of gestures. Depending on the volume of collected data, modeling approaches range from support vector machines (SVM), trees, \eg \cite{georgi_recognizing_2015,iravantchi_beamband_2019}, to deep learning models, \eg \cite{hu_fingertrak_2020,yeo_opisthenar_2019}.


Our approach trains a high-performance model from large volumes of data (collected from our user study). Moreover, we take this process one step further by extending that model's ability to recognize novel gestures (\eg customized by a new user) with a few samples. We discuss the details of our approach in Sections~\ref{sub:design:pretrained} and \ref{sub:design:newhead}.

\subsection{Gesture Customization}
\label{sub:related_work:customization}
The advantages of supporting customized, user-defined gestures include but are not limited to greater memorability~\cite{nacenta_memorability_2013}, higher interaction efficiency~\cite{ouyang_bootstrapping_2012}, and better accessibility for people with physical disabilities~\cite{anthony_analyzing_2013}. 
Prior work has explored and summarized customized gesture sets through user elicitation studies (\eg \cite{wobbrock_user-defined_2009,ruiz_user-defined_2011,piumsomboon_user-defined_2013}), and others built tools that facilitate the creation of new customized gestures (\eg \cite{yang_magic_2012,bau_octopocus_2008,oh_challenges_2013}).

To enable a favorable experience for end-users, gesture customization systems need to support a nimble yet effective data collection process.
The HCI field has examined several approaches for supporting gesture customization by demonstration~\cite{dey_cappella_2004,lu_gesture_2012}, including rule-based approaches ~\cite{avrahami_guided_2001,doring_gestural_2011}, and tiered computational methods~\cite{lou_personalized_2017,anthony_lightweight_2010, mckenna_comparison_2004,ouyang_bootstrapping_2012}.
Related to our work, uWave~\cite{liu_uwave_2009} stored templates of accelerometer signals for new gestures, and used DTW to compare against incoming data streams.
Bigdelou \etal~\cite{bigdelou_adaptive_2012} applied Laplacian Eigenmaps and kernel regression on arm-worn IMU signals, while
Mezari \etal~\cite{mezari_easily_2018} leveraged fast Fourier transforms (FFT), symbolic aggregate approximation, and simple distance metrics to recognize new gestures.

Although these systems require minimal training data, they often only work with hand gestures that involve significant hand motion, where the IMU signals have high variance. As we will show in Section~\ref{sec:study_algorithm}, traditional methods perform poorly when applied to complex, fine-grained gestures.

\subsection{Related Machine Learning Techniques}
\label{sub:related_work:machine_learning_techniques}
Our work intersects with several ML-based approaches.
These include \textit{transfer learning}~\cite{gao_knowledge_2008,pan_survey_2010}, a method that focuses on applying knowledge gained from solving one task to another related task, and \textit{incremental learning}~\cite{wu_automated_2020,masana_class-incremental_2021,polikar_learn_2001}, an approach that accommodates new data to continuously extend a model's knowledge without full retraining. Specifically, our method belongs to a subcategory of transfer learning: solving new tasks (\ie new gestures) in the same domain (\ie hand gesture recognition)~\cite{pan_survey_2010}.
Likewise, the goal of learning to recognize new gestures with few samples fits within the \textit{few-shot learning problem}~\cite{wang_generalizing_2020}. A number of techniques have been proposed to address few-shot learning, including metric learning~\cite{kaya_deep_2019}, meta-learning~\cite{gong_metasense_2019,finn_model-agnostic_2017}, and multi-task learning~\cite{caruana_multitask_1997,zhang_survey_2021}.
Within the gesture recognition domain, few-shot learning is performed on camera data~\cite{stewart_online_2020,wu_one_2012} or EMG signals~\cite{rahimian_few-shot_2021}.
However, previous research neglected the problem of extending an existing model (for pre-existing classes) to include new classes.

Our framework is a variant of dynamic few-shot learning~\cite{gidaris_dynamic_2018,tao_few-shot_2020}, where the goal is to train a model that can learn base categories, while dynamically recognizing novel categories from only a few training examples. Our approach is a combination of transfer learning, incremental learning and few-shot learning methods, which we describe in the next two sections.

\section{Pre-trained Gesture Model}
\label{sub:design:pretrained}

We designed a system that integrates transfer learning, class incremental learning, and few-shot learning for gesture customization. Figure~\ref{fig:overview_framework} visualizes the structure of the framework. In this section, we describe our pre-trained model in detail, and we present our gesture customization model in Section~\ref{sub:design:newhead}.

\subsection{Data Collection}
\label{subsub:design:pretrained:data}
We sought to build a five-class classifier that can recognize four dynamic hand gestures (Clench, Double Clench, Pinch, and Double Pinch, see Figure~\ref{fig:gesture_set}) and one non-gesture case (\ie negative gesture).  To ensure robustness, we made significant effort to build a large-scale and diverse hand gesture dataset. Table~\ref{tab:pretrain_data_collection} offers a comprehensive summary.

\renewcommand{\arraystretch}{1.3}
\begin{table*}[b]
\centering
\resizebox{0.8\textwidth}{!}{
\begin{tabular}{ccl}
\hline
\hline
                                         & \textbf{Factor}                  & \multicolumn{1}{c}{\textbf{Information}}                                                                   \\
\hline                                         
\multirow{3}{*}{Demographics}            & Total Number                  & 512 Participants                                                           \\
                                         & Self-identified Gender                  & Female 133, Male 378, Non-binary 1                                                                \\
                                         & Age                     & Min 21, Max 63, Mean 33.1$\pm$10.5                                                                \\
                                         & Hand Habits             & Right handed (442), left-handed (70)                                                               \\ \cline{1-3}
\multirow{4}{*}{Gesture Data}  & Body Posture            & $\begin{array}{@{}l@{}}\hbox{Sitting upright (N=224), Sitting and leaning back (N=149),}\\\hbox{Standing (N=58), Lying down (N=21), and others (N=60)}\end{array}$\\
                                         & Eye-hand Angles         & $\begin{array}{@{}l@{}}\hbox{Abdomen-level (N=193), Chest level (N=38),}\\\hbox{Head-level (N=203), and others (N=78)}\end{array}$                                     \\                               \\
                                         & Motion                  & Static (472), Walking (40)  \\   
                                         & Gesture Variation       & Regular (499), Intentionally slower/faster/weaker/stronger (13)                                   \\
                                         & Contexts                & Gesures only (500), Gestures inserted with regular chores (12)                                    \\ \cline{1-3}
\multirow{2}{*}{Negative Data} & In-lab Daily Activities & Walking, using mobile phones, typing (500)                                                        \\
                                         & Targeted Negative Data  & A wide range of behaviors that involve fine-grained hand movement (12) \\
\hline
\hline
\end{tabular}
}
\vspace{0.1cm}
\caption{Data Collection Information to Build The Pre-trained Model}
\label{tab:pretrain_data_collection}
\Description{
Tables of the data collection information. The first group of rows presents demographics, which is broken down into four factors: total number, self-identified gender, age, and hand habits. The second group of rows presents gesture data, which is broken down into five factors: body posture, eye-hand angle,s motion, gesture variation, and contexts. The third group of rows presents negative data, which is broken down into two factors: in-lab daily activities and targeted negative data.
}
\end{table*}
\renewcommand{\arraystretch}{1.0}

\subsubsection{Participants and Apparatus}
\review{Leveraging a user-experiment-platform with a large user repository,} we recruited 512 participants (133 self-identified female, 378 male, 1 non-binary) with a wide coverage of age range (min=21, max=63, mean=33.1$\pm$10.5). Majority of users were right-handed (N=442, left-handed=70).
We used Apple Watch Series 5 and 6 for data collection, with IMU sensors sampled at 800~Hz max, ultimately downsampled to 100~Hz during training \footnote{\review{We collected raw data with an overly high sampling rate, 800~Hz, to maximize our dataset ability. However, during the model training, we found that 100~Hz already suffices. Thus, in the rest of the paper, our framework only uses 100~Hz data.}}.
Participants wore the watch on their non-dominant hand during the data collection.
Data was first stored on the watch and then uploaded to a server for processing and model training.
\review{The user study received institutional IRB approval.}

\subsubsection{Gesture Data}
We asked participants to follow instructions on the watch throughout a session. In each round, a gesture would appear on the screen, and we asked participants to perform that gesture 10 times, each time following a three-second countdown. Gesture order was randomized and each participant performed at least three rounds of data for each gesture.
Throughout this process, a phone was placed directly above the user's hand, and we recorded video to serve as additional ground truth for annotation purposes.

We also considered other relevant factors: body posture, hand-eye angle, motion, gesture variation, and activity.
We randomly picked a subset of participants to perform gestures in different body postures, such as sitting upright (N=224), sitting and leaning back (N=149), standing (N=58), and lying down (N=21).
For different hand-eye angles, participants were either asked to put down their hand to abdomen-level (N=193), chest level (N=38), or head-level (N=203).
We also asked a few participants (N=40) to walk while performing gestures.
Lastly, we asked a small fraction of participants (N=13) to perform gestures at different speed and intensities (\ie slower-faster, weaker-stronger).
Some participants (N=12) performed light everyday tasks (\eg typing or wrist twisting) while occasionally performing a gesture.

\subsubsection{Negative Data}
In addition to positive examples, we also asked participants to perform negative (\ie non-gesture) examples. In this round, we asked participants to perform normal indoor daily activities, such as walking, phone browsing, and typing (among others).
Moreover, we asked a small group of participants (N=12) to perform a wide range of behaviors that involved fine-grained hand movement, including tapping on the watch/other surfaces, scratching head/hands, using mouse/keyboard, playing pens, brushing teeth, shaving, washing hands/utensil, showering, driving, using juicer/vacuum cleaners, playing video games, opening bottles, and biking. These negative sessions were not video-recorded.

\subsubsection{Annotation}
Each data collection study lasted 30 to 60 minutes.
We synchronized videos with our IMU signals and annotated the start and end times of each gesture performance. After annotation, we collected approximately 110,000 gesture samples (30 hours), and 60 hours of negative samples.
The average duration of single gestures (\ie Clench and Pinch) is around 550ms, while the average duration of double gestures (\ie Double Clench/Pinch) is around 800ms.

\begin{figure*}[t]
    \centering
    \includegraphics[width=0.8\textwidth]{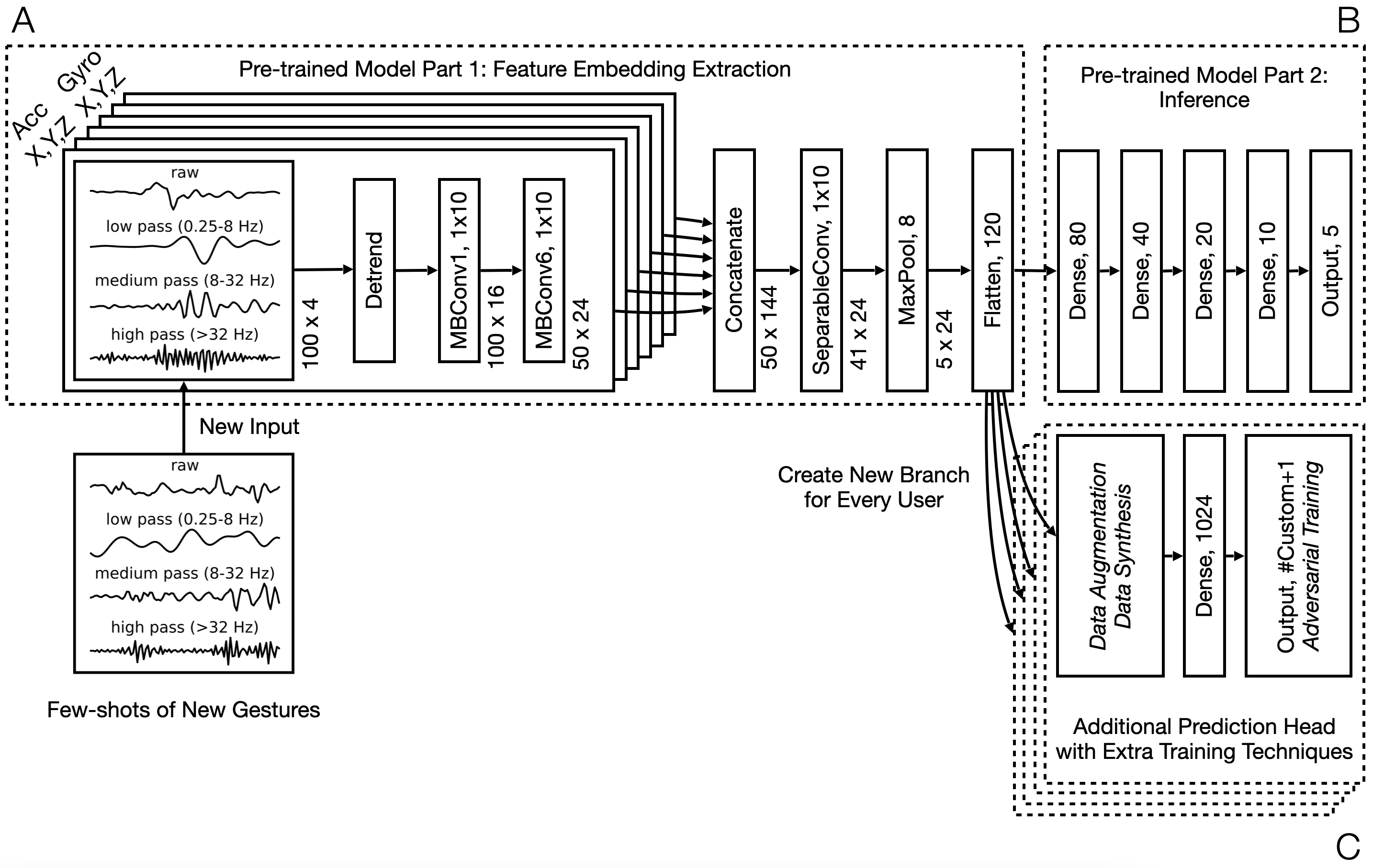}
    \caption{Hand Gesture Customization Framework. The upper region shows the two parts of a pre-trained gesture recognition model: feature embedding extraction (A) and inference (B). The lower region shows the additional prediction head (C) that trained on new inputs of the new gestures. Both the pre-trained model and the new model shared the same embedding layers.}
    \label{fig:overview_framework}
    \Description{
The architecture of the deep learning model. The upper region shows the two parts of a pre-trained gesture recognition model: feature embedding extraction and inference. The feature embedding extraction part takes raw signals as input, feeds them through a few convolutional neural network blocks, and passes them to the inference part. The inference part consists of a few fully connected layers.
The lower region shows the customization architecture that takes the output of the feature embedding extraction layers and feeds it into data augmentation and synthesis block, followed by a fully connected layer, followed by an output layer with adversarial training.
    }
\end{figure*}

\subsection{Model Architecture}
\label{subsub:design:pretrained:model}
The raw input of the model is a one-second, six degree-of-freedom accel-gyro signal (three axes for accelerometer, and three axes for gyroscope) sampled at 100~Hz.
Each channel is preprocessed using three Butterworth bandpass filters (0.22-8 Hz, 8-32 Hz, 32 Hz) using cascaded second-order sections, leading to a $100\times4$ input size for every channel.

We adopt the concept of EfficientNet~\cite{tan_efficientnet_2019} to balance the number of trainable parameters and model performance. Specifically, for each input channel, we employed two inverted residual blocks~\cite{sandler_mobilenetv2_2018} (\ie MBConv in Figure~\ref{fig:overview_framework}) to process the signals. We then concatenate the output of the six channels, and add one more separable convolution layer~\cite{chollet_xception_2017} to capture concatenated information with low computational cost, followed by a max-pooling layer and a flatten layer.
We mark these layers as the feature embedding extraction part of the pre-trained model (Figure~\ref{fig:overview_framework}a), whose output is a one-dimension vector with a vector length of 120.

The latter half of the pre-trained model consists of a stack of five fully connected layers with sizes 80, 40, 20, 10, and 5. We insert a batch normalization layer~\cite{ioffe_batch_2015} and a dropout layer ($p=0.5$)~\cite{gal_dropout_2016} between every two fully connected layers to improve model generalizability. The output of the final layers correspond to the confidence of the five classes. We use cross-entropy as the loss function, and Adam optimizer during the training. The entire model has 106k total parameters, a suitable size for on-device inference. 

\subsection{Model Training and Performance}
\label{subsub:design:pretrained:performance}
After data collection, we processed each data sequence with a sliding window mechanism, with the window size as 1 second (same as the input of the model), and a step size of 0.125 sec (simulating an 8Hz prediction frequency).
We then randomly split 50\%, 10\%, 40\% of the dataset into training, validation, and testing sets. It is worth noting that data splitting was conducted based on participants so that the evaluation outcomes are cross-user results.
We trained our model for 200 epochs, with a 0.1 exponential learning rate decay every 50 epochs. The epoch with the best results on the validation set is saved and evaluated on the testing set.

\begin{figure}
    \centering
    \begin{subfigure}[b]{.45\columnwidth}
        \centering
        \includegraphics[width=1\columnwidth]{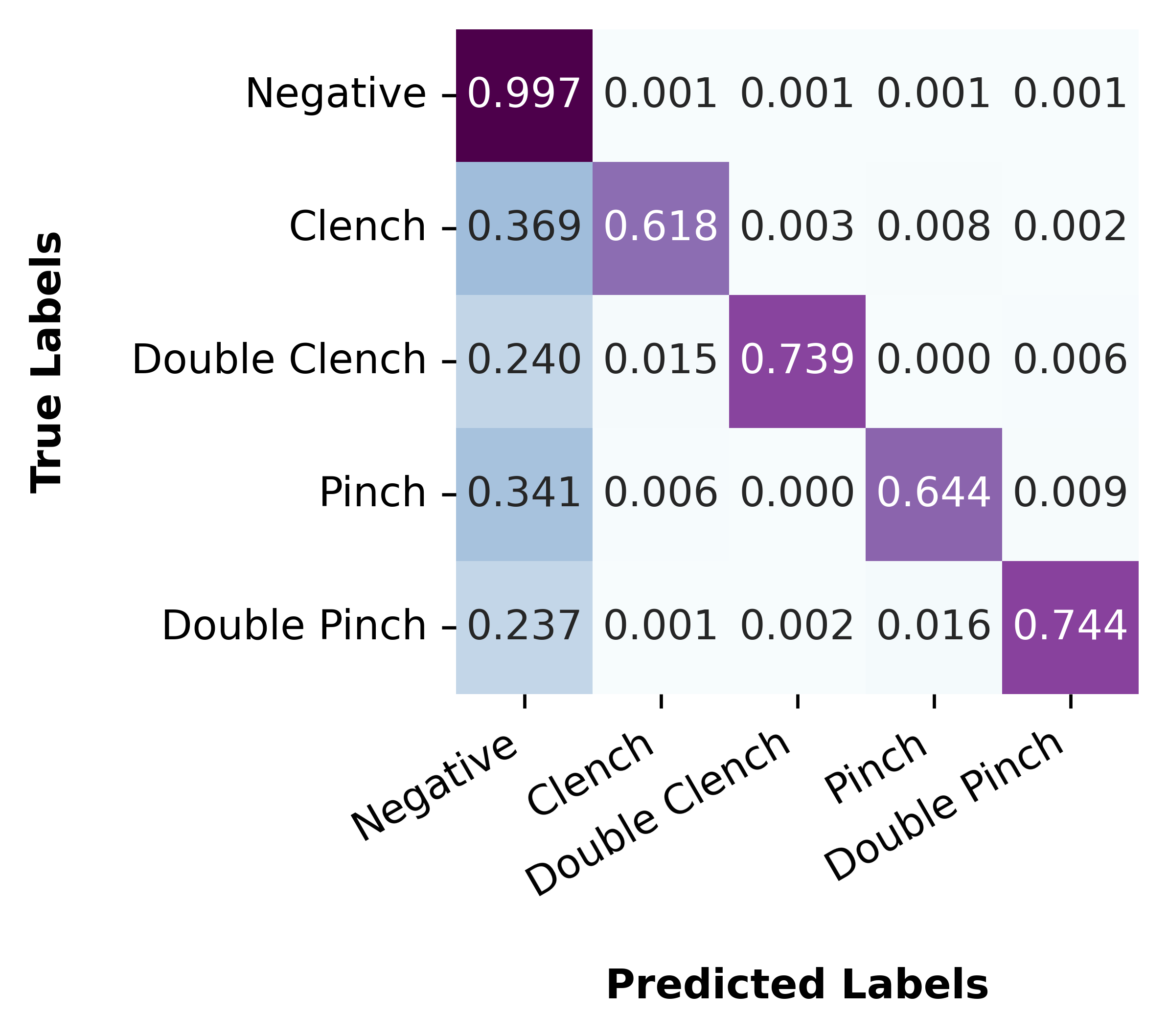}
        \caption{Window-level Matrix}
        \label{subfig:cfmtx_pretrain:window}
    \end{subfigure}
    \quad
    \begin{subfigure}[b]{.45\columnwidth}
        \centering
        \includegraphics[width=1\columnwidth]{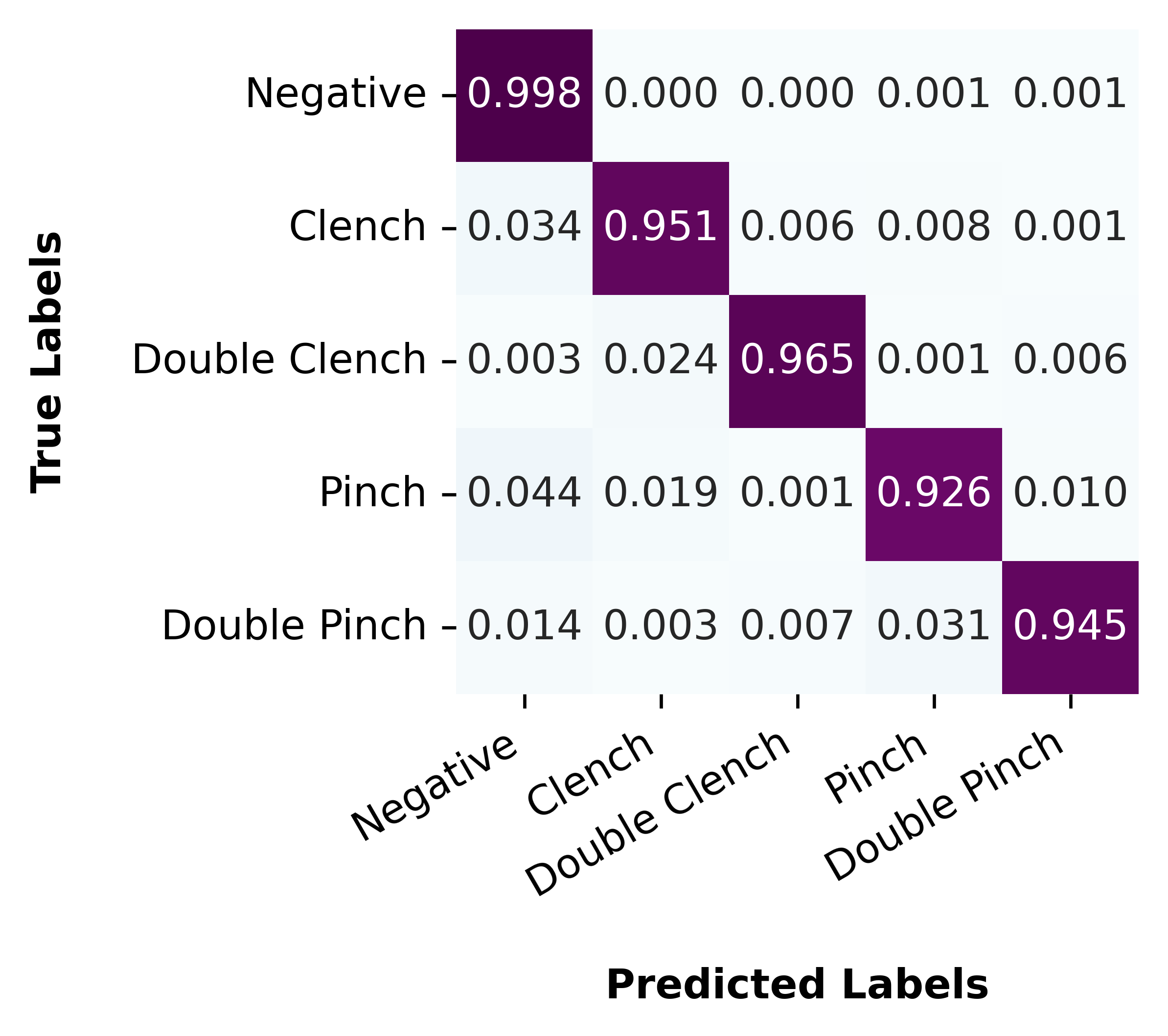}
        \caption{Gesture-level Matrix}
        \label{subfig:cfmtx_pretrain:gesture}
    \end{subfigure}
    \caption{Prediction Results as Confusion Matrices of The Pre-trained Model. The window-level prediction (a) has an average accuracy of 74.8\% and an F1 score of 82.7\%. The gesture-level aggregation (b) significantly improves the results, with an accuracy of 95.7\% and an F1 score of 95.8\%.}
    \label{fig:cfmtx_pretrain}
    \Description{
Two confusion matrices of the pre-trained model recognition results, one at the window level and one at the gesture level.
(a) A 5 by 5 confusion matrix of the window level. Five columns from the left to the right are negative, clench, double clench, pinch, and double pinch. The diagonal numbers of the matrix is 0.996, 0.618, 0.739, 0.644, and 0.744. The four gestures are most easily confused with the negative. The first column numbers except negative itself are 0.369, 0.240, 0.341, and 0.237 for the four gestures, respectively. Other numbers are all below 0.02.
(b) A 5 by 5 confusion matrix of the gesture level. Five columns from the left to the right are negative, clench, double clench, pinch, and double pinch. The diagonal numbers of the matrix is 0.998, 0.951, 0.963, 0.926, and 0.945. The first column numbers except negative itself are 0.034, 0.003, 0.044, and 0.014 for the four gestures, respectively. Other numbers are all below 0.025.
    }
\end{figure}

\subsubsection{Window-level Prediction Performance}
We first investigate the direct outcome of the prediction, which is at the window level.
The results show an average accuracy of 74.8\%,  a precision of 93.6\%, a recall of 74.8\%, and an F1 score of 82.7\%. Figure~\ref{subfig:cfmtx_pretrain:window} visualizes the confusion matrix of the window-level prediction on the testing set.

The confusion matrix shows that there is little confusion among the four gestures, and that the majority of the misclassification comes from the model not recognizing some windows where a gesture actually happens, leading to false negatives.
The window-level prediction only focused on every single 1-sec window. However, in real-time scenarios, a gesture is comprised of multiple windows. Therefore,  we need to aggregate our window-level predictions to obtain our  gesture-level predictions.

\subsubsection{Gesture-level Prediction Performance}
The aggregation involves a few hyper-parameters. For the four gestures, we need to decide how many consecutive prediction windows are required before the model predicts a gesture. For negative data, we also need to decide how many consecutive windows with non-negative predictions are required before the model triggers a ``false positive''.
We performed grid search on these hyper-parameters using our validation set. Our final consecutive window length thresholds are 3, 4, 3, 4 for Clench, Double Clench, Pinch, and Double Pinch, respectively.

Aggregation significantly improves our recognition performance, with an average accuracy of 95.7\%, a precision of 95.8\%, a recall of 95.7\%, and an F1 score of 95.8\%. Moreover, the gesture-level false positive rate is 0.6 times per hour.
Figure~\ref{subfig:cfmtx_pretrain:gesture} presents the confusion matrix of the gesture-level results.
These results indicate that our model can accurately recognize gestures on new users' data and is highly robust to negative data.

\section{Gesture Customization}
\label{sub:design:newhead}
Having a model that can recognize four gestures and works across users with robust performance, we now describe our gesture customization framework.

\subsection{Customization Architecture}
\label{subsub:design:newhead:model}
\begin{figure}
    \centering
    \includegraphics[width=\columnwidth]{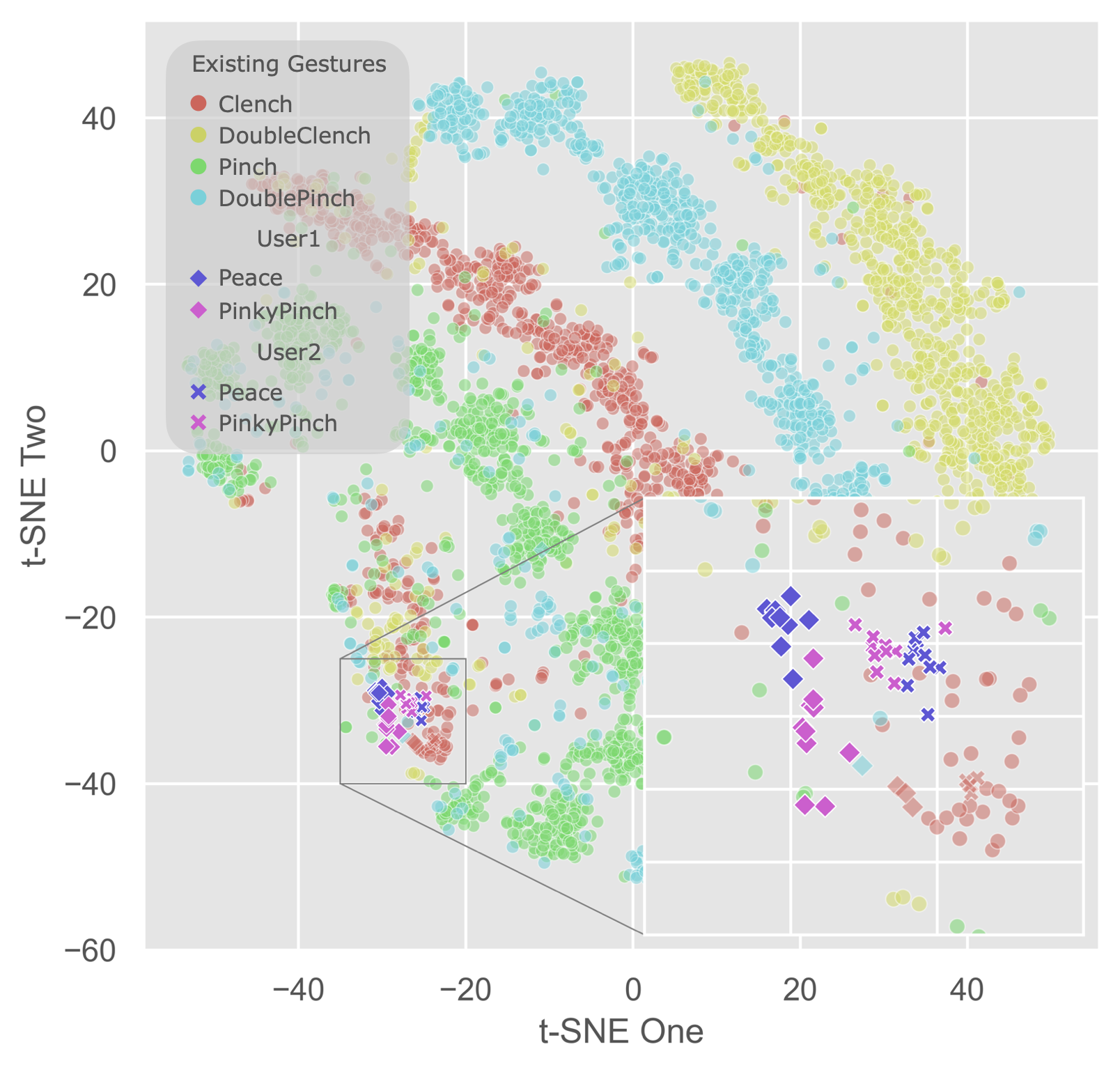}
    \caption{2-D t-SNE Visualization of The Feature Embedding Vectors of Different Gestures. The same color indicates the same gesture. Data collected for building the pre-trained model is plotted in circle, while new users' customized gesture data are plotted in cross and diamond.
    The zoom-in area suggests that even for the same customized gesture, different users' data may have distinct clusters.}
    \label{fig:embedding_tsne}
    \Description{
A two-dimensional t-SNE plot. The data points of the four existing gestures form four big clusters. A small amount of data of these four gestures, together with the data of new gestures from new users form another small cluster.
Zooming in the small cluster shows that 1) different users' data form different smaller clusters for each user. 2) within each users' own data, different gestures further form clusters for each gesture.
    }
\end{figure}

Our framework integrates transfer learning, class incremental learning, and few-shot learning.
After building the pre-trained model with good performance, we create a new branch after the feature embedding extraction layers as the additional prediction head (Figure~\ref{fig:overview_framework}a and Figure~\ref{fig:overview_framework}c).
Note that each user will have their own particular branch, even when two users want to add the same customized gesture.
To better understand this, Figure~\ref{fig:embedding_tsne} visualizes the 2-dimensional t-distributed stochastic neighbor embedding (t-SNE) plot with a subset of the four gestures' data collected in Section~\ref{subsub:design:pretrained:data}, as well as two new users' data with both of them performing two customized gestures (Peace and PinkyPinch, see Figure~\ref{fig:gesture_set}) and two old gestures (Clench and Pinch).

There are a few observations. First, most of the four pre-existing gestures form clear clusters, which reflects the high accuracy of the pre-trained model.
More importantly, the two users' data has some interesting patterns: while both users did the same gestures, each user's own data form a cluster and the two users' clusters are not close to each other, especially for the gesture Peace. This indicates high between-user variance and low within-user variance, \ie a user can do a gesture in a relatively consistent way, while two users may do it quite differently.
Besides, the decision boundaries to distinguish the two new gestures (Peace \vs PinkyPinch) are also different between the two users.
These observations further support building a customized prediction head for each user's customization gestures.

We employ a simple, light-weighted two-layer fully connected network, the first as a feature processing layer and the second as the output layer. The first layer uses Leaky ReLU ($\alpha=0.3$)~\cite{nair_rectified_2010} as the activation function and has a L2 kernel regularizer ($\lambda =5e{-5}$)~\cite{hecht-nielsen_theory_1992} and a dropout layer ($p=0.5$)~\cite{gal_dropout_2016} to reduce overfitting.
The last layer uses Softmax activation that corresponds to the prediction confidence of the final classes. The number of the classes is equal to the number of customized gestures plus one more for the negative case.
Therefore, when users create their first customized gesture, the additional prediction head is trained as a binary classifier. When a second gesture is added, a new three-class prediction head is trained from scratch, \etc
Since the prediction head is light-weighted, the training process is fast.

In real-time, the new prediction head works together with the pre-trained model to recognize gestures. The two models recognize distinctive gestures and are both robust to negative data.
If both models predict a gesture, the one with the highest confidence would be the final prediction.

Our framework leverages the first half of the pre-trained model as the feature extractor and transfers it to new gesture recognition tasks. By training the new prediction head for incremental classes, the performance of the existing gestures is not impacted, addressing the \textit{forgetting old} problem.
Then, we tackle the few-shot challenge with a series of data processing techniques.

\subsection{Maximizing Few Shots}
\label{subsub:design:newhead:techniques}
\begin{figure*}
    \centering
    \includegraphics[width=1\textwidth]{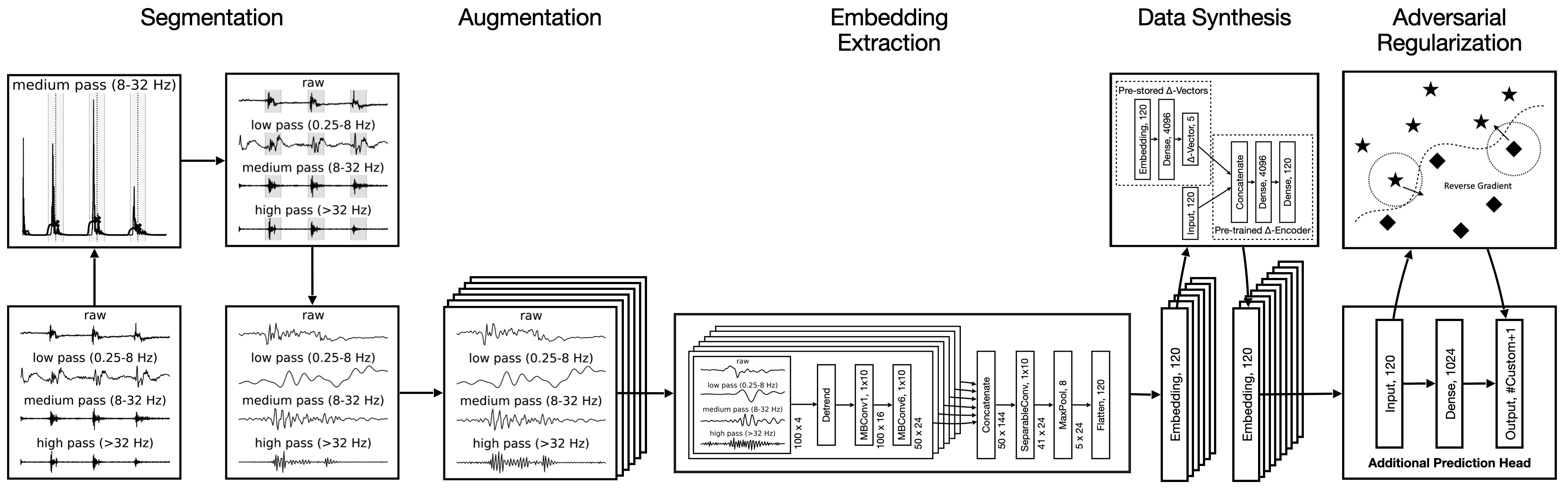}
    \caption{The Data Processing Pipeline of Training a Gesture Customization Prediction Head. It starts with a short data sequence (3 shots in this example) recorded by an end-user and goes through data segmentation, augmentation, and synthesis before the training. The training process is enhanced by adversarial regularization.}
    \label{fig:fewshot_processing_pipeline}
    \Description{
The overall data process pipeline for our framework. It contains the following five parts from the left to the right: segmentation, data augmentation, embedding extraction, data synthesis, and adversarial regularization.
    }
\end{figure*}

No matter how much we simplify the model, training a model with less than 10 samples is challenging. It can easily fall into the overfitting problem. It is also hard to prevent false-positives as the model does not have enough ``positive'' samples (\ie gesture samples) to learn from.
We use a series of techniques to make the most out of the small amount of data provided by users (see Figure~\ref{fig:fewshot_processing_pipeline}).

\subsubsection{Data Segmentation}
Before actual data processing, it is worth noting that there is no readily available training sample.
When end-users record data of their new customized gestures, they can either do gestures consecutively in a row (similar to the data collection process in Section~\ref{subsub:design:pretrained:data}), or follow some instructions to do one gesture at a time and repeat several times, depending on the interaction design. In either way, it is unrealistic to ask users to provide the exact start and end timestamp of the gesture.
Therefore, we need to segment the signal sequence to obtain data samples.

We take the output signals of the middle bandpass filters (8-32 Hz) as this is robust to noisy arm movement, and use a peak detection algorithm to identify potential moments of performing hand gestures.
Specifically, we calculate the sum of the magnitude of both filtered accelerometer and gyroscope signals, and apply a 1-sec absolute moving average to smooth the data. We then use a simple peak detection method that finds local maxima by comparing neighboring values (with distance threshold as 1 sec).
A peak is ignored if it is lower than the overall average of signal magnitude. If any time reference is available (\eg a countdown mechanism), we can further filter peaks according to the reference.
We take a 1-sec window centered at these potential peaks, and feed them into the feature extraction part of the pre-trained model.
We then compute a Euclidean distance matrix of the normalized embedding vectors and remove outliers (threshold empirically set as 0.8).
In such a way, we can segment out pronounced, repetitive hand movement periods that correspond to the target gestures.

Once the peaks are determined, we take a 1.5-sec window centered at each final peak to ensure that a gesture is fully covered by the window. Our data augmentation techniques are applied to these windows.

\subsubsection{Data Augmentation}
After data segmentation, we use several data augmentation techniques to generate a larger number of samples.
\review{Three time series data augmentation techniques~\cite{iwana_empirical_2021}, with all seven combinations ($2^3-1$), are used to generate positive samples, enlarging the data size by eight times}: 1) zooming, to simulate different gesture speed, randomly chosen from $\times0.9$ to $\times1$; 2) scaling, to simulate different gesture strength, with the scaling factor $s \sim \mathcal{N}(1,0.2^2), s \in [0,2]$, and 3) time-warping, to simulate gesture temporal variance, with 2 interpolation knots and warping randomness $w \sim \mathcal{N}(1,0.05^2), w \in [0,2]$.

Moreover, we also employ three augmentation techniques to generate gesture data that is marked as negative~\cite{xu_listen2cough_2021}: 1) cutting out by masking a random 0.5 sec of signals by zero; 2) reversing signals, and 3) shuffling by slicing signals into 0.1-sec pieces and generating a random permutation. These augmentations are often used in other ML tasks to augment positive data, but we mark the data augmented by these techniques as negative to ensure our model only recognizes valid gestures. We also applied the seven positive augmentation techniques on these negative data to generate more negative samples.

After data augmentation, we take a 1-sec sliding window on these 1.5-sec windows to generate samples to be fed into the pre-trained model. The step size is set as 0.1 sec, leading to five input samples from each 1.5-sec window.
In addition, the data collected in Section~\ref{subsub:design:pretrained:data} are all added as negative data to improve the robustness of the model against noisy movement.

\subsubsection{Data Synthesis}
Although the data augmentation can generate signals with larger variance from the data recorded by end-users, these augmented data may not be close to the actual gesture variance introduced by natural human behavior.
Therefore, we further synthesize more data from both the raw signals and the augmented signals that simulate the natural motion variance. 
Specifically, we train a $\Delta$-encoder~\cite{schwartz_-encoder_2018}, a self-supervised encoder-decoder model that can capture the difference between two samples (\ie $\Delta$) belonging to the same gesture, and use it to synthesize more new gesture samples.

A $\Delta$-encoder is trained as follows: it takes two samples (sampleInput and sampleRef) from the same class as the input, feeds sampleInput through a few neural network layers to be a very small embedding called $\Delta$-vector (similar to a typical Autoencoder~\cite{goodfellow_deep_2016}), and then use the $\Delta$-vector and the sampleRef to reconstruct sampleInput. The intuition comes from the fact that the size of $\Delta$-vector is so small that it focuses on capturing the difference between sampleInput and sampleRef, which is then used to rebuild sampleInput with sampleRef as the reference~\cite{schwartz_-encoder_2018}.
After the $\Delta$-encoder is trained, it can take another sample from the new class as a new sampleRef, and generate a new sample of the same class with a $\Delta$-vector. This $\Delta$-vector can either be obtained via feeding any existing sample from other classes through the encoder, or randomly generated.

In our case, we use the data of the four pre-existing gestures to train a $\Delta$-encoder.
During the training, we randomly draw two samples from \textit{the same gesture and the same user} to ensure that the model captures the within-user variance instead of the between-user variance.
We use the feature embeddings of length 120 as the input and the output of the $\Delta$-encoder to save computation cost. Our structure of our $\Delta$-encoder is fairly simple. Both the encoder and decoder have one hidden layer with a size of 4096 and uses Leaky ReLU ($\alpha=0.3$) as the activation function. The size of $\Delta$-vector is set as 5.
Using the training set from the four gestures, the model is trained with 200 epochs and has a 0.5 exponential decay on the learning rate every 30 epochs. The epoch with the best results on the validation set is saved. We also calculate and save the $\Delta$-vectors from the four gestures' testing set, which will be used to generate new samples.

\review{In real-time, when the customized gestures' data come in and go through the augmentation stage, we use the $\Delta$-encoder to generate extra samples of the customized gestures that contain more natural gesture variance, enlarging both augmented positive and negative data by 10 times.
}

\subsubsection{Adversarial Training Regularization}
After the data augmentation and data synthesis, we obtain a large amount of data with appropriate variance to train the prediction head.
To further improve the robustness of the model, we adopt the practice of adversarial training regularization~\cite{goodfellow_explaining_2015,ma_towards_2018} when learning the model.

The main idea of adversarial regularization is to train a model with adversarially-perturbed data (perturbed towards the decision boundary or inverse gradient decent so that the training process becomes harder) in addition to the original training data.
It can prevent the model from overfitting and classify the data points close to the boundary more robustly.
In Figure~\ref{fig:embedding_tsne}, the two customized gestures' data from the same user tend to be blended with the existing four gestures near the boundary. The adversarial regularization can help to enhance classification performance, especially for the purpose of reducing false-positive.
We set both the adversarial regularization loss weight and the reverse gradient step size as 0.2.

Through a series of data segmentation, data augmentation, data synthesis, and adversarial training, we can learn a robust prediction head for each new user that can accurately recognize their customized gestures with a low false-positive rate.
Figure~\ref{fig:fewshot_processing_pipeline} visualizes the whole training procedure of the prediction head.

\subsection{Interactive Feedback}
\label{subsub:design:newhead:ux_roadmap}

\begin{figure*}[t]
    \centering
    \includegraphics[width=1\textwidth]{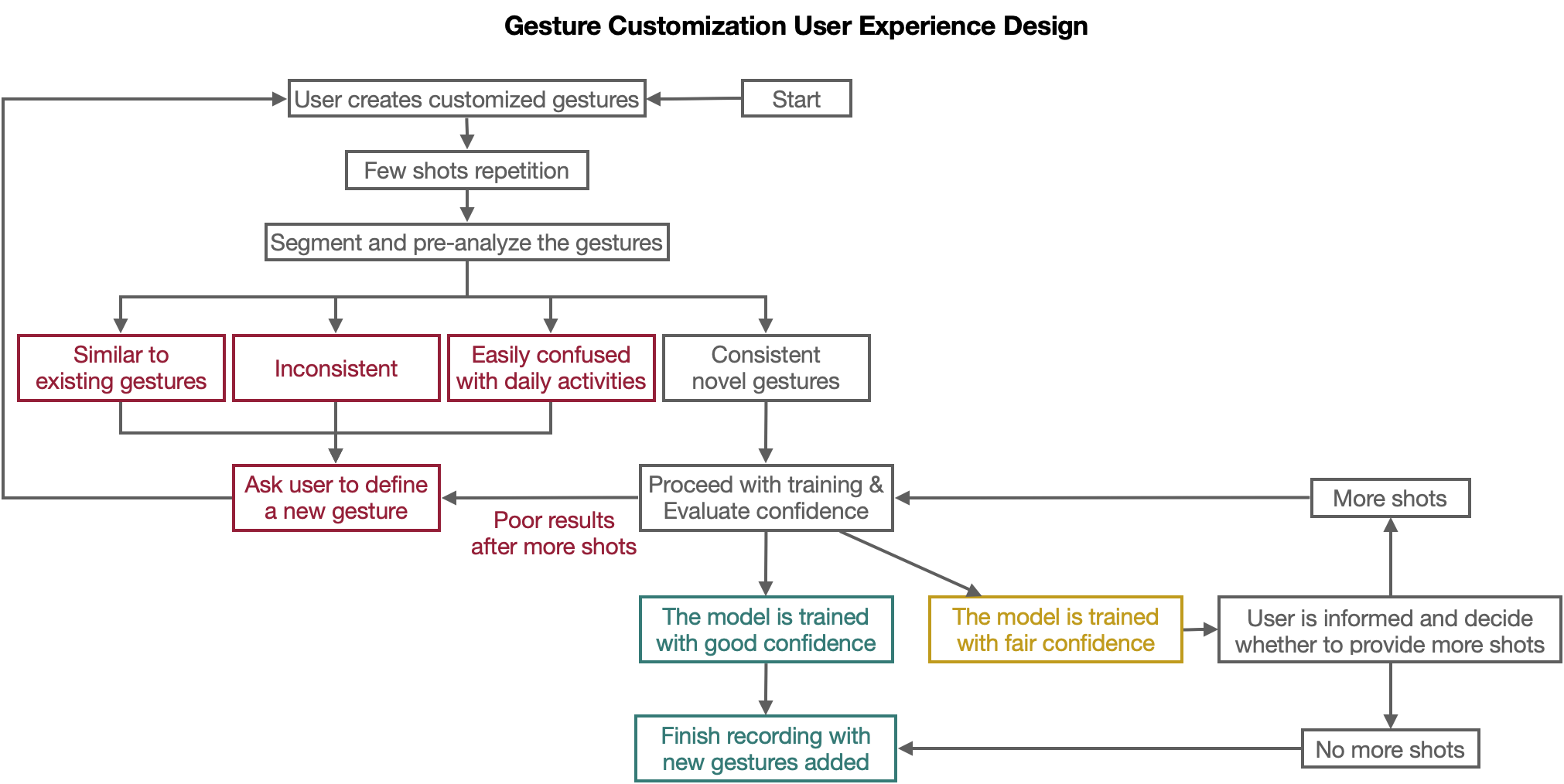}
    \caption{Gesture Customization User Experience Design. If a new gesture is similar to existing gestures, or performed inconsistently, or close to daily activities, the framework will provide corresponding feedback to users and ask them to define another gesture. Moreover, if a new gesture is novel and performed consistently, but the model is trained with fair performance, the framework will offer users to choose finishing or collecting a few more samples. Such a feedback can help users to better understand the process and design gestures.}
    \label{fig:ux_roadmap}
    \Description{
A flow chart showing the gesture customization experience design. The process starts with "User create customized gestures", "Few shots repetition", and "Segment and pre-analyze the gestures".
Then, depending on the collected samples, it will move to one of the following four situations: "Similar to existing gestures", "Inconsistent", "Easily confused with daily activities", and "Consistent novel gestures". For the first three conditions, it will go to "Ask users to define a new gesture" and loop back to the block "User create customized gestures". For the last condition with consistent and novel gestures, it will move to "Proceed with training \& Evaluate the confidence", which will have two outcomes: 1) "The model is trained with good confidence", so that it will "Finish the recording with new gestures added", or 2) "The model is trained with fair confidence", and then "Users are informed and decide if they want to provide more shots". It shows two choices: picking "No more shots" will lead to "Finish the recording with new gestures added". Picking "More shots" will ask users to provide more shots and go back to "Proceed with training \& Evaluate the confidence". If the model has good results, it will "Finish the recording with new gestures added". If the model still has poor results, it will lead to "Ask users to define a new gesture" and go back to the initial stage.
    }
\end{figure*}

We take a few prerequisites into account to design the user experience.
When building the prediction head, we have two implicit assumptions: 1) Each customized gesture is unique and distinguishes from existing gestures; 2) The sequence provided by an end-user does contain valid, consistent gesture repetitions.
Moreover, to avoid frequent false-positive triggers, a gesture should not be close to ordinary daily actions, such as shaking (common in teeth-brushing, washing, and scratching) and slow waving (easily involved in driving or greeting).

Therefore, our framework should not simply accept any incoming data provided by end-users. Instead, it needs to be sanity checked to ensure the gesture are reliable~\cite{oh_challenges_2013}.
We design the overall gesture customization user experience for our framework, as shown in Figure~\ref{fig:ux_roadmap}.


When the data is recorded and segmented, we examine whether it belongs to any of the following three situations and provide real-time feedback to users to help users better understand the process and design gestures~\cite{oh_challenges_2013}:
\begin{s_itemize}
\item \textit{Similar to existing gestures.} We feed the segmented data into the pre-trained model and the additional prediction head (if it exists). If either model predicts the majority of the segments to be one of the existing gestures, it indicates that the new gesture is close to previous gestures.
\item \textit{Inconsistent.} During the segmentation, we check the euclidean distance matrix of potential gesture repetitions and filter out those that are far from the rest of the repetitions (see Section~\ref{subsub:design:newhead:techniques}). After the filtering, if the number of repetitions left is less than the expected number (\eg 3 when the framework requires a three-shot recording), it means that users did not do the gesture in a consistent way.
\item \textit{Easily confused with daily activities.} To find whether the new gesture is close to common daily behaviors, we leverage the negative data collected in Section~\ref{subsub:design:pretrained:data}. We use the pre-trained model to extract the embeddings of the negative data in the testing set (sliced in 1-sec pieces), and apply Hierarchical Density-Based Spatial Clustering of Applications with Noise (HDBSCAN)~\cite{campello_hierarchical_2015} to automatically cluster the data. HDBSCAN is a variant of DBSCAN~\cite{ester_density-based_1996} that can adapt different distance thresholds based on the cluster density, obviating the necessity of setting this hyperparameter. We use the euclidean distance as the metric, and set the minimal cluster size as 3. HDBSCAN identifies 2,500 clusters. We calculate and save the center of these clusters and use them as the representation of common daily activities. After the new gesture data is segmented, we compute a distance matrix between the gesture data and these cluster centers, and find each gesture sample's closest center. If the majority of the gesture data are close to at least one of these centers (threshold empirically set as 0.4), it means that the new gesture is close to common daily activities.
\end{s_itemize}

When a customized gesture is novel and performed consistently, the framework will proceed and go through a series of \review{data augmentation, data synthesis, and adversarial training.}
After the training, we synthesize extra gesture data and use them as a testing set.
If the testing accuracy is good enough (set as 80.0\%), the process is completed and the model can recognize the new gesture.
When the accuracy is below the threshold, the framework will inform users of the accuracy value and let them decide either to whether to continue data recording and re-train the model with more data. If the model still does not perform well on the testing set after the second data collection, it will ask users to define a new gesture.

\section{Evaluation}
\label{sec:study_algorithm}
We evaluate our framework from two aspects.
In this section, we focus on the algorithmic perspective and measure the model performance on various new gestures.
In the next section, we assess it from the usability perspective and test the real-time system via a user study.

\subsection{Data Collection}
\label{sub:study_algorithm:data}
\review{We conducted a user study to collect data from 16 gestures (four existing gestures in Section~\ref{sub:design:pretrained} and 12 new gestures) to train customized gesture recognition for each individual.}

\subsubsection{Gesture Design}
\begin{figure*}
    \centering
    \includegraphics[width=1\textwidth]{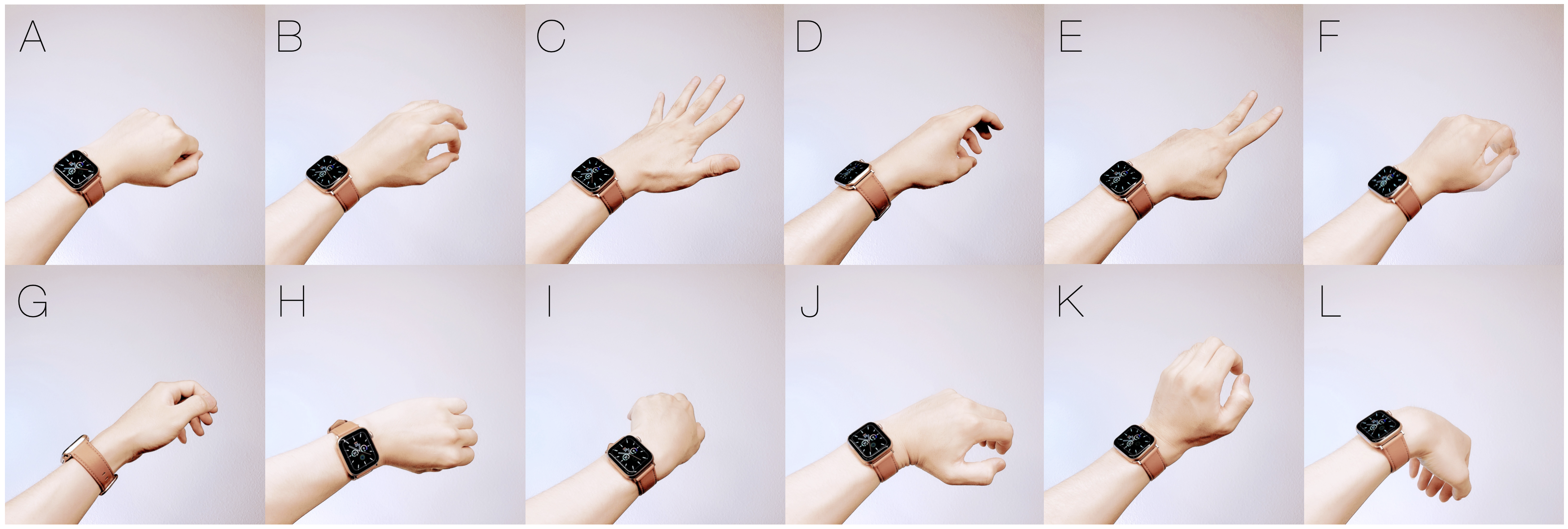}
    \caption{Dynamic Hand Gesture Set Involved in the Study: (A) Clench and DoubleClench, (B) Pinch and DoublePinch, (C) Spread and DoubleSpread, (D) PinkyPinch and DoublePinkyPinch, (E) Peace, (F) Slide, (G) RotateOut, (H) RotateIn, (I) DeviateOut, (J) DeviateIn, (K) Extend, (L) Flex. All gestures start from a neutral, relax hand pose, and return back to the neutral pose at the end. The four gestures of (A) (B) are supported by the pre-trained model. It is worth noting that this gesture set is only for the purpose of evaluation. Our framework can work with a much more wide range of gestures as long as they are not close to existing gestures or common daily activities.}
    \label{fig:gesture_set}
    \Description{
12 pictures of a hand wearing a watch, each performing different geutres: (A) Clench (also indicate DoubleClench), (B) Pinch (also indicate DoublePinch), (C) Spread (also indicate DoubleSpread), (D) PinkyPinch (also indicate DoublePinkyPinch), (E) Peace, (F) Slide, (G) RotateOut, (H) RotateIn, (I) DeviateOut, (J) DeviateIn, (K) Extend, (L) Flex.
}
\end{figure*}

To evaluate the performance of our framework, we refer to the taxonomy of dynamic hand gestures~\cite{choi_taxonomy_2014} and \review{choose a set of new gestures (in addition to the four supported by the pre-trained model) that covers a wide range of movement.
Other than the existing four gestures (Clench, Double Clench, Pinch, Double Pinch), we pick a set of 12 new gestures} that are representative of different wrist/hand/finger movement patterns~\cite{choi_taxonomy_2014}: Spread and Double Spread have opposite finger movement (opened \vs closed) against Clench and Double Clench, PinkyPinch and Double PinkyPinch use a different finger than Pinch, and Peace has two fingers opened and three fingers closed; Slide has opposite motion between the thumb and the index finger; TwistOut/In, RotateOut/In and Extend/Flex involve wrist movement in different ways.
Figure~\ref{fig:gesture_set} visualizes the 16 gestures (12 new + 4 existing). All gestures start from the neutral pose and return back to the neutral pose.

It is worth noting that the main purpose of this gesture set is to evaluate the framework's performance. The actual gesture set that can be supported by our framework goes beyond these 12 gestures.

\subsubsection{Participants and Apparatus}
20 participants (4 self-identified female, 16 male, Age = 29.8$\pm$4.9) volunteered to participate in the data collection study. 3 participants are left-handed.
We employed the Apple Watch Series 6 for data collection, with IMU sensors sampled at 100 Hz.
All participants wore the smartwatch on their non-dominant hands.

\subsubsection{Design and Procedure}
Participants went through multiple sessions for data collection. Each session is similar to the study in Section~\ref{subsub:design:pretrained:data}, where participants saw the gesture name on the watch screen, followed a 3-sec countdown to perform the gesture, and repeated five times.

Every participant started with one session for each of the existing four gestures as a warm-up stage. Then, they had five data collection sessions. In each session, they performed each of the 12 new gestures 5 times. A Latin-square design was used to counterbalance the order effect.
Participants took a 30-sec break between gestures and a 2-min break between each session to reduce fatigue.
\review{Moreover, to simulate the actual use case of taking the watch on and off over time, participants were asked to take off and put on the watch during the break to vary the watch position on the wrist, with a variation within 5 cm.}
The study was around 30 to 40 minutes.
Overall, for each participant, we collected 5 repetitions per existing gesture, 25 repetitions (5 sessions $\times$ 5 repetitions) per new gesture.

\subsection{Model Performance}
\label{sub:study_algorithm:results}

We followed the procedure depicted in Figure~\ref{fig:fewshot_processing_pipeline} to process the data. For each user, we randomly picked two sessions (\ie 10-shot maximum) as the training set, one session as the validation set, and the remaining two as the testing set.
We repeated three times and computed the average performance.
We also evaluated the robustness of the model against noise by applying it to the negative testing set and measuring the false-positive rate.

Note that during the testing, we use a sliding window mechanism on the whole sequence to simulate a real-time use case. Thus, similar to Section~\ref{subsub:design:pretrained:performance}, the results can be evaluated at both window-level and gesture-level.
For the gesture-level prediction, we set a uniform consecutive window length threshold as 5.

\review{
In the rest of the section, we first evaluated the recognition performance on the new gestures (Section~\ref{subsub:study_algorithm:results:new}). We then evaluated whether the recognition on the existing gestures were impacted after introducing new prediction heads (Section~\ref{subsub:study_algorithm:results:existing}). We further combined existing and new gestures, and evaluated their overall performance (Section~\ref{subsub:study_algorithm:results:all}). Finally, we compared our framework with a range of baseline techniques (Section~\ref{subsub:study_algorithm:results:baseline}). 
}

\subsubsection{Recognizing New Gestures}
\label{subsub:study_algorithm:results:new}
We investigated two factors that have important design implications: the number of shots used for training and the number of new gestures that the model is trained to recognize.
For the first factor, we went through different numbers of training samples from the original training set (from 1 shot to 10 shots) to train the model. 
For the second factor, given the number of new gestures, we went through all possible combinations of the gestures, from one new gesture to four new gestures.
In total, we trained and evaluated 475,800 models (10 shot numbers $\times$ 3 repetition $\times$ $\sum_{n=1}^{4}{12 \choose n}$=793 gesture combinations  $\times$ 20 participants).

\begin{figure}
    \centering
    \includegraphics[width=\columnwidth]{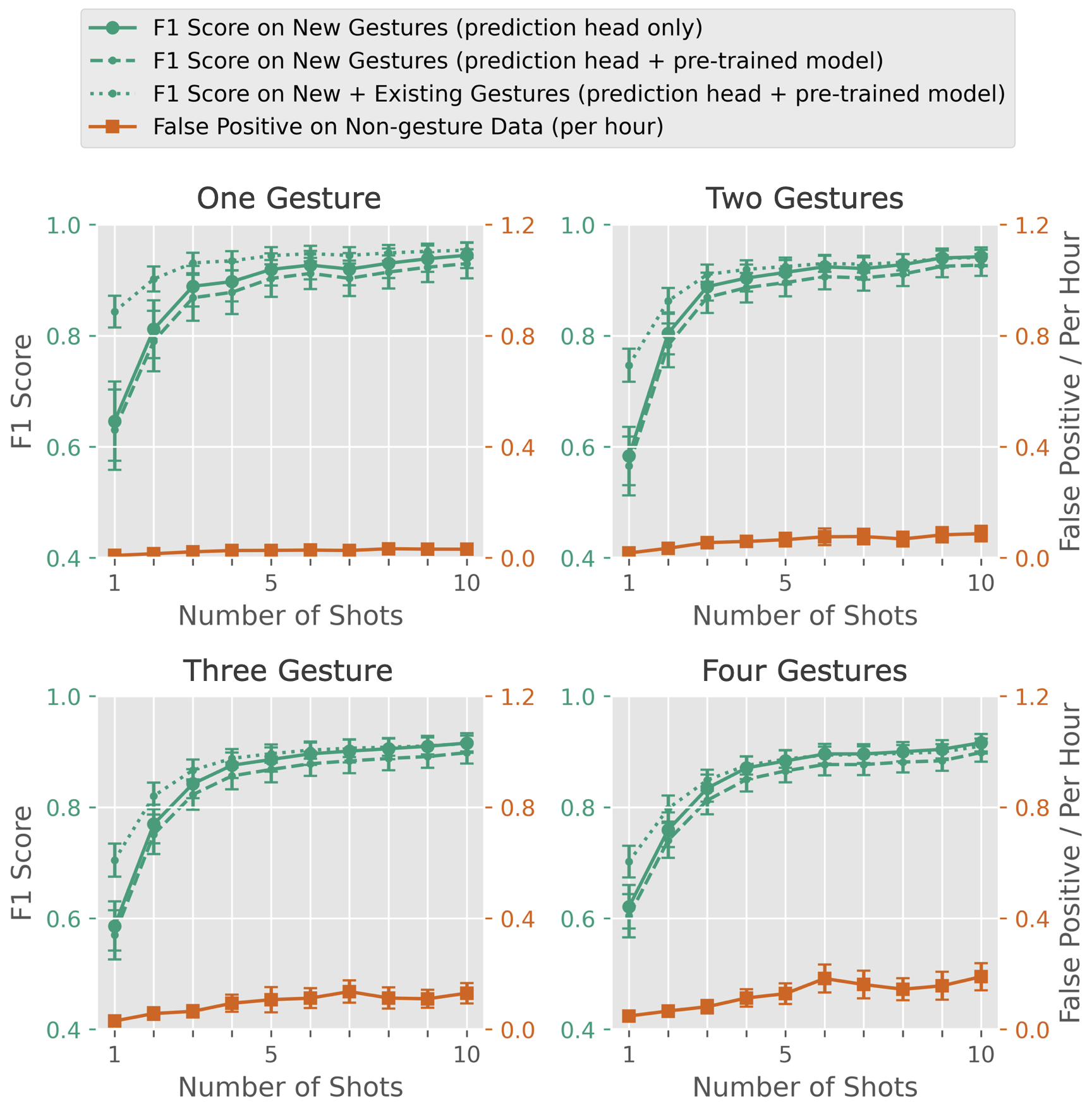}
    \caption{Prediction Performance with Different Number of Shots and Gestures. The accuracy and F1 score results correspond to the left y-axis, while the false positive rate results correspond to the right y-axis. Error bars indicate the standard error of the mean.}
    \label{fig:fewshots_results}
    \Description{
Four plots show the prediction performance results on adding one, two, three, four gestures. In each figure, X-axis is the number of shots from 1 to 10. The Y-axis has twin axes, the one on the left shows the F1 score, and the one on the right shows the false positive rate per hour.
There are three F1 score lines in each plot, corresponding to F1 Score on New Gestures (prediction head only), F1 Score on New Gestures (prediction head + pre-trained model), and F1 Score on New + Existing Gestures (prediction head + pre-trained model). And there is one false positive line in each plot.
Overall, the higher the number of shots is, the higher the F1 scores are, while the false positive rate keeps low. As adding more gestures, the task becomes more challenging and but the performance only has small drops.
}
\end{figure}


\textbf{Prediction Head Evaluation.}
We first evaluate the performance of the prediction head (the solid lines in Figure~\ref{fig:fewshots_results}).
When using only one shot to add a new gesture  (\ie users perform the gesture just one time), our framework can achieve an average gesture-level accuracy of 55.3\% and an F1 score of 64.6\%.
The more shots the model has, the better the recognition performance.
With three shots of a new gesture, our framework can achieve an average accuracy of 83.1\% and an F1 score of 88.9\%.
The performance further increases to 87.2\% and 92.0\% with five shots, and 91.0\% and 94.5\% when using ten shots.
Meanwhile, the models are robust to noisy data, with an average false positive rate af 0.02 times per hour when evaluated on daily activity non-gesture data.

\begin{figure}[t]
  \centering
    \includegraphics[width=0.8\columnwidth]{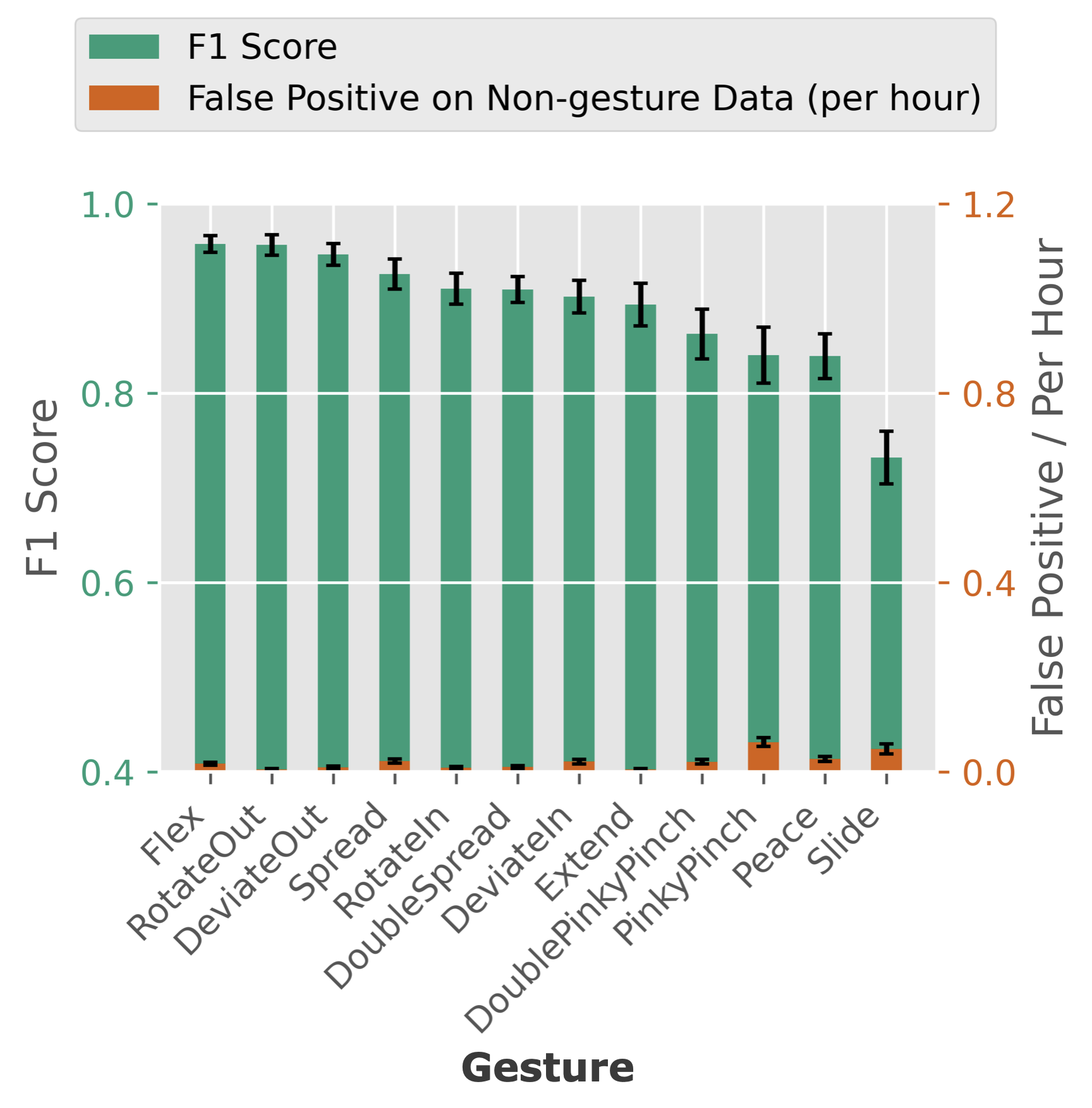}
    \caption{Prediction Performance of One New Gesture.}
    \label{subfig:fewshots_results_1_2_gestures:one}
  \Description{
Results evaluation of adding one gesture: Barplot of the performance of every single gesture. The Y-axis is the same as the plots in Fig 8. The X-axis shows the 12 gestures ranked by their F1 scores, with the highest on the left, and the lowest on the right. The gestures from the left to right are: Flex, RotateOut, DeviateOut, Spread, RotateIn, DoubleSpread, DeviateIn, Extend, DoublePinkyPinch, PinkyPinch, Peace, and Slide.
 }
\end{figure}
\begin{figure}[b]
  \centering
    \includegraphics[width=0.85\columnwidth]{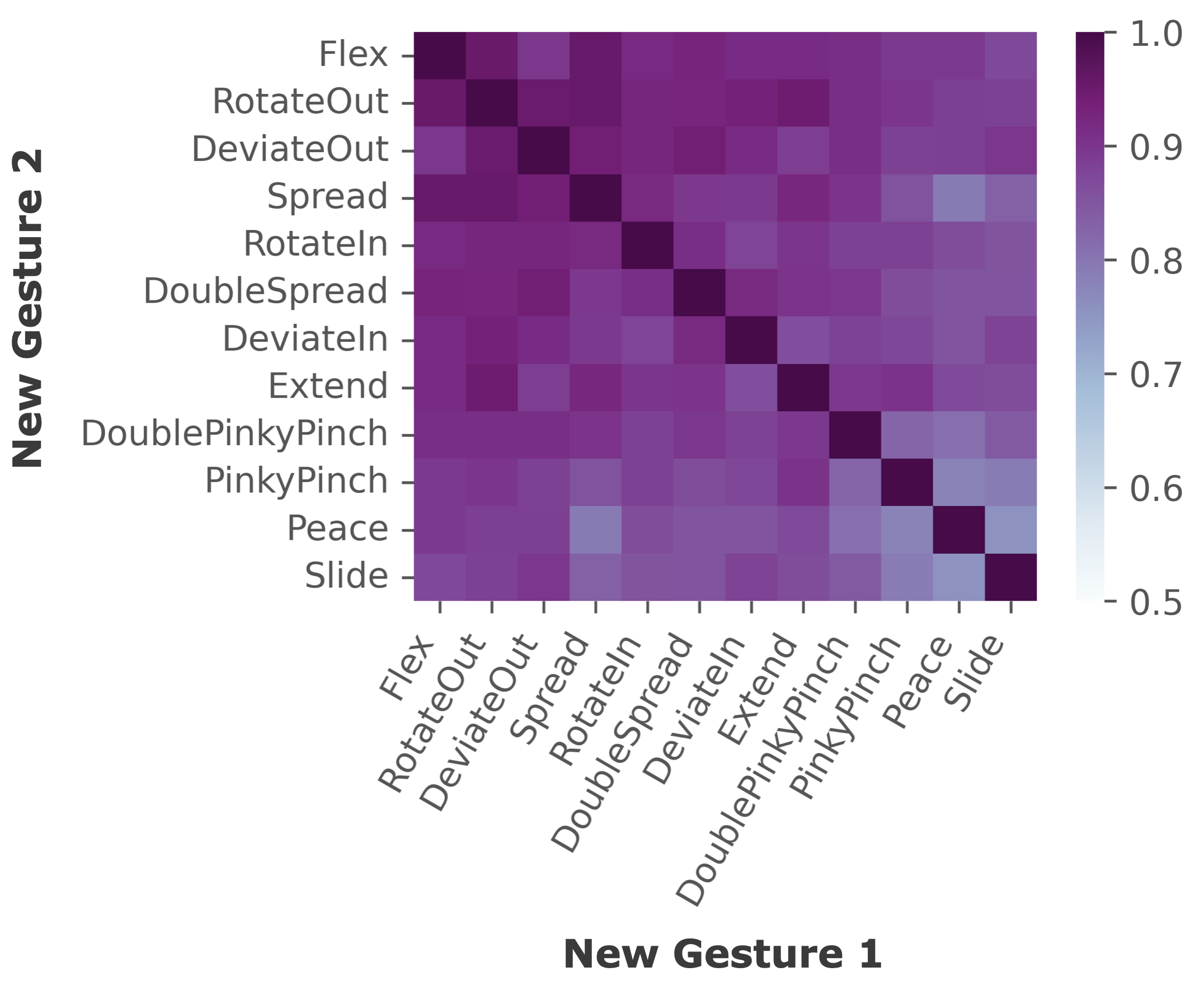}
    \caption{\review{Prediction Performance of Two New Gestures. Higher F1 score indicates less confusion between the two newly added gestures.}}
    \label{subfig:fewshots_results_1_2_gestures:two}
  \Description{
Results evaluation of adding two gestures: A 12 by 12 F1 score matrix when every pair of two gestures are added as new gestures. The column and the raw order is the same as Fig 9.a. Overall, most pair of gestures have high F1 scores. The last three gestures PinkyPinch, Peace, and Slide have relatively lower scores around 75\%.
}
\end{figure}

Supporting more than one gesture is more challenging, but our framework achieves an average accuracy of 83.3\% and an F1 score of 88.8\% with three shots of two new gestures.
When adding three new gestures, our method has an accuracy of 77.7\% and an F1 score of 84.2\%. 
For four gestures, our method still has an accuracy of 77.2\% and an F1 score of 83.4\%.
\review{More new gestures also lead to a slightly higher false positive rate, and we observe the same trend as more number of shots are included for training. This can be explained by the fact that the increased variety of the positive samples raises the difficulty of the classification task.} But our models can maintain the false positive rate as low as 0.06 or 0.12 times per hour when adding two or four gestures.
The evaluation results show promising potential of the framework.

Moreover, we investigate the individual performance of each gesture.
Figure~\ref{subfig:fewshots_results_1_2_gestures:one} reveals that the majority of the 12 gestures have good performance.
Using three shots, 7 out of 12 gestures have F1 scores at least 90\%. Spread, RotateOut, and Flex have F1 scores higher than 95\%.
In contrast, Slide has relatively lower performance.
The difference can be explained by the fact that the sliding gesture has larger variation, or is hard for the accelerometer/gyroscope to capture the motion.

To evaluate how these new gestures are confused against each other, we also look into the confusion between each pair of gestures when both are added as new gestures. Figure~\ref{subfig:fewshots_results_1_2_gestures:two} shows that Slide and Peace are relatively more easily confused with other gestures. 

\textbf{Combining Prediction Head and Pre-trained Model.}
After the prediction head is trained and applied in real-time, it works with the pre-trained model together for recognition.
Therefore, we also evaluate the performance on the new gestures when combining the two models, as shown by the dashed lines in Figure~\ref{fig:fewshots_results}.
The results are very close to those tested solely on the prediction head, with a minimal performance drop of 1.7\% on F1 scores.
This indicates that the two models have a distinguishing focus on different gestures (\ie new gestures \vs existing four gestures) and barely confuse each other.


\subsubsection{Recognizing Existing Gestures}
\label{subsub:study_algorithm:results:existing}
In addition to evaluating the performance of our framework on new gestures' data and negative data, it is also important to measure how much the additional prediction head influences the recognition performance on the existing four gestures.
Both the average accuracy and F1 score achieve 97.5\% and 97.7\% when applying the combined model on participants' data of the existing four gestures.
This shows that the recognition outcomes of the original four gestures are not impacted by the additional prediction head.



\subsubsection{Recognizing All Gestures}
\label{subsub:study_algorithm:results:all}
The real-time system in actual usage can recognize both new gestures and existing gestures.
Therefore, we also evaluate the two models on all gestures, as shown by the dotted lines in Figure~\ref{fig:fewshots_results}.
When using one shot to add a new gesture, our framework can achieve an average F1 score of 84.3\% on the five gestures. With three, five, and ten shots, the F1 score increases to 93.1\%, 94.4\%, and 95.4\%.
When adding two, three, or four gestures with three shots, the average F1 scores achieve 91.0\%, 86.6\% or 84.9\% on the whole six, seven, or eight gestures.
\review{We summarize detailed results in Appendix Tab.~\ref{tab:fewshots_results}}

The results from Section~\ref{subsub:study_algorithm:results:new} to Section~\ref{subsub:study_algorithm:results:all} suggest that our framework can include new gestures and achieve good performance with only a few shots, without degrading the recognition accuracy on existing gestures.

\subsubsection{Comparing to Other Methods}
\label{subsub:study_algorithm:results:baseline}
We also compare our framework against a few other methods. Some of them are traditional computational methods, while some of them are deep-learning-based:
\begin{s_itemize}
\item \textit{DTW}. In some prior work~\cite{liu_uwave_2009}, DTW can be used to recognize new hand waving gestures with only one sample as the template. We re-implement the algorithm in uWave~\cite{liu_uwave_2009} and test it using our datasets.
\item \textit{Traditional ML models}. As deep learning methods are usually data-hungry, an alternative solution is to use off-the-shelf traditional ML models to lower the data requirement. We test both SVM and random forest as they are commonly used on wearable gesture recognition systems (\eg \cite{zhang_tomo_2015,georgi_recognizing_2015,iravantchi_beamband_2019}). The input for these traditional models are the feature embeddings from the pre-trained model.
\item \textit{Fine-tuning on the pre-trained model}. This method is one of the common transfer-learning-based solutions. Specifically, we remove the final layer of the pre-trained model and add a new layer with more output nodes (five original classes plus the number of new gestures). We copy the weights from the old layers for the old five nodes, and randomly initiates the weights for the new nodes. Then, we fine-tuning the model using new data.
\item \textit{Ablation study}. In addition to other methods, we also remove one of the data augmentation, data synthesis, and adversarial regularization methods to evaluate each of their individual contribution to the final results.
\end{s_itemize}

\renewcommand{\arraystretch}{1.3}
\begin{table}[]
\centering
\resizebox{1\columnwidth}{!}{
\begin{tabular}[t]{c|ccc|ccc}
\hline
\hline
\multirow{2}{*}{\textbf{Methods}}  & \multicolumn{3}{c}{\textbf{Window-level}}      & \multicolumn{3}{|c}{\textbf{Gesture-level}}       \\ \cline{2-7}
 & \textbf{acc}    & \textbf{F1}    & \textbf{FP Rate}    & \textbf{acc}  & \textbf{F1}    & \textbf{FP \#/Hr}      \\ \hline
DTW & 0.485 & 0.355 & 0.515 & 0.552 & 0.597 & 0.457 \\
SVM & 0.928 & 0.737 & 0.000 & 0.712 & 0.796 & 0.021 \\
Random Forest & 0.895 & 0.686 & 0.000 & 0.686 & 0.763 & 0.000 \\
Fine-tuning & 0.915 & 0.516 & 0.039 & 0.448 & 0.511 & 6.175 \\ \cline{1-7} 
\textbf{w/o Augmentation} & 0.898 & 0.497 & 0.000 & 0.244 & 0.327 & 0.000 \\
\textbf{w/o Synthesis}  & 0.933 & 0.784 & 0.001 & 0.819 & 0.879 & 0.034 \\
\textbf{w/o Adv Regularization}  & 0.922 & 0.742 & 0.002 & 0.792 & 0.855 & 0.106 \\
\textbf{Full Pipeline} & \textbf{0.935} & \textbf{0.790} & \textbf{0.001} & \textbf{0.833} & \textbf{0.888} & 0.055 \\
\hline
\hline
\end{tabular}
}
\vspace{0.1cm}
\caption{Results Comparison between Our Framework and Other Methods. All training and testing use three shots and two new gestures to ensure consistency.}
\label{tab:fewshots_baseline}
\Description{
Comparison between our techniques and baseline methods. There are eight rows in the table, each representing a method. The first four are baselines: DTW, SVM, random forest, and fine-tuning. The last four are our framework with different ablations: our framework without augmentation, our framework without data synthesis, our framework without adversarial regularization, and our completed framework.
Each row has the results on both window-level and gesture-level. Each level includes accuracy, F1 score, and false-positive rate (for window-level) or count (for gesture-level).
}
\end{table}
\renewcommand{\arraystretch}{1.0}

To make a fair comparison, we use three shots and two gestures for consistency, and negative data is available in all methods.
Table~\ref{tab:fewshots_baseline} presents both the window-level and gesture-level results.
Our method significantly outperforms the traditional methods and the fine-tuning method by at least 12.1\% on accuracy and 9.2\% on F1 score.
Moreover, the ablation study results show that each of the techniques helps improve the model performance.

\section{Usability}
\label{sec:study_usability}
\begin{figure*}[t]
    \centering
    \includegraphics[width=1\textwidth]{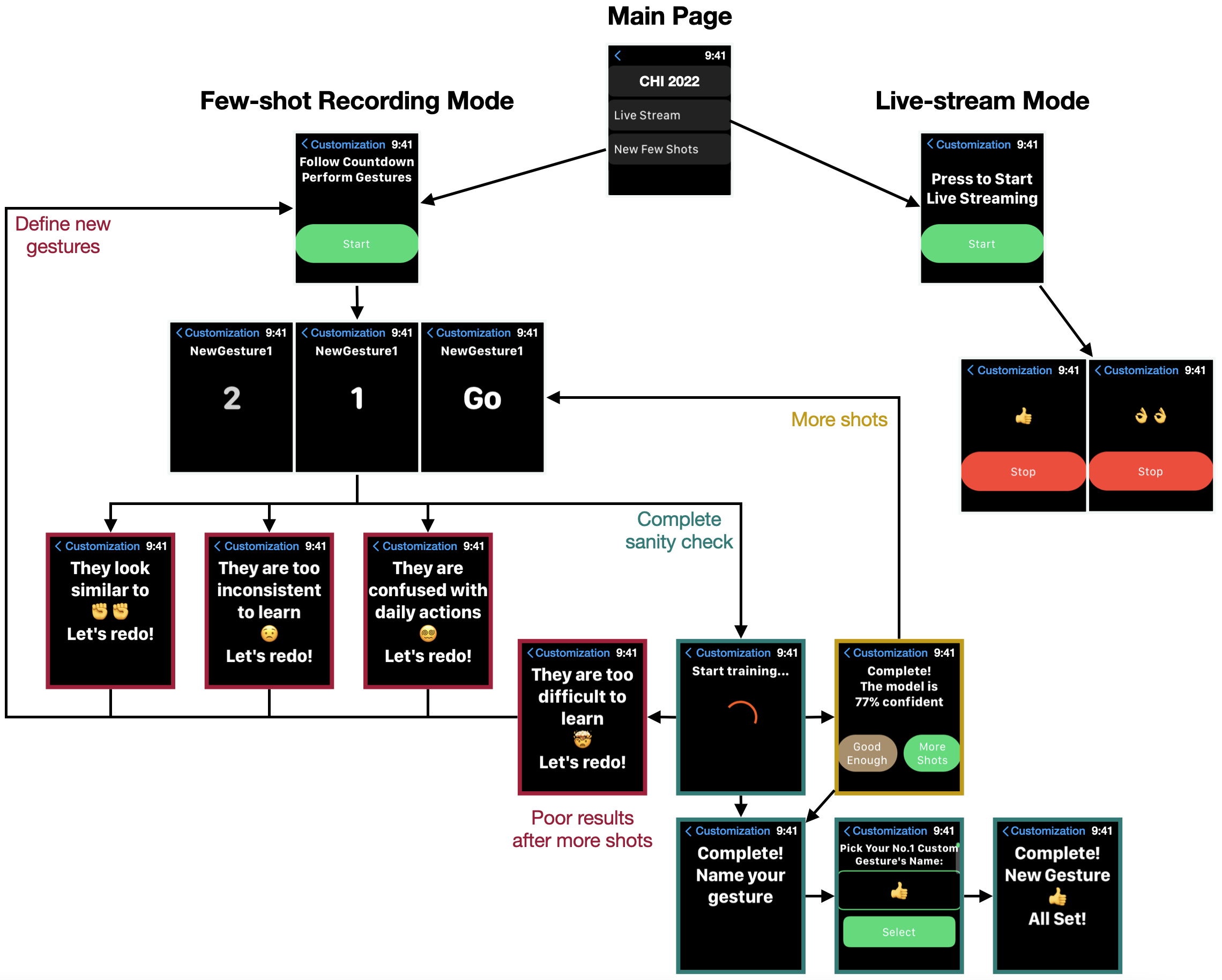}
    \caption{Real-Time Gesture Customization System Watch Interface Design. The few-shot recording mode is consistent with the user experience design in Figure~\ref{fig:ux_roadmap}.}
    \label{fig:interface_roadmap}
    \Description{
A series of watch interface screenshots for our real-time system. It is divided into two groups: live-stream mode and few-shot recording mode. The live-stream mode shows simple interfaces when a gesture is detected in live steam. Screenshots in the few-shot recording mode are consistent with Fig 6 - the experience design.
    }
\end{figure*}

Finally, we implemented a real-time system based on our framework, and evaluated the system via a user study.
Figure~\ref{fig:interface_roadmap} shows the watch interface that corresponds to the user experience roadmap.
Not only do we evaluate the real-time recognition performance, more importantly, we also measure the system usability and collect users' feedback.

\subsection{Participants and Apparatus}
\label{sub:study_usability:users}
We invited the same set of users in Section~\ref{sec:study_algorithm} for the usability evaluation.
Apple Series S6 was used as the apparatus for the usability study, worn on participants' non-dominant hands.
For prototyping purposes, the watch streamed the data to a MacBook Pro laptop. The laptop did the model training and gesture recognition, and sent the results back to the watch for real-time interaction.

\subsection{Design and Procedure}
\label{sub:study_usability:design}
We used a three-shot version of the system for the evaluation. Participants went through the following stages after a brief introduction of the system and the interface:
\begin{s_enumerate}
\item Participants first tried the recognition system in the live-stream mode with the existing four gestures to get themselves familiarized with the system.
\item Participants recorded a pre-defined new gesture RotateOut three times to add the gesture, and tested it in the live-stream mode, together with the four gestures (five in total).
\item Participants were asked to create one or more customized gestures themselves and record the gesture. After adding it successfully, they tested it in the live-stream mode, together with the five existing gestures.
\item Participants were told to use the system freely for 5 to 10 minutes, after which they filled in the System Usability Scale (SUS)~\cite{bangor_empirical_2008} and Task Load Index (NASA-TLX) questionnaire~\cite{hart_development_1988}, and had a semi-structured interview about their experience.
\end{s_enumerate}
During the study, participants could ask the experimenter anytime if they have any questions. The experimenter noted down all successful or unsuccessful recognition results. The study lasted about 30 minutes.

\subsection{Results}
\label{sub:study_usability:results}
Overall, many participants were excited about the system. We summarize \review{the recognition results, customization procedure, and subjective feedback.}

\subsubsection{New Gestures Recognition Performance}
\review{
The real-time performance of the system is similar to the offline results in Section~\ref{sub:study_algorithm:results}.
The average recognition accuracy and F1 score on the four existing gestures are 96.7\% and 98.1\%. For new gestures, the overall average accuracy and F1 score are 91.1\% and 95.1\%, respectively.
The customized gestures defined by participants were diverse. Examples include snapping fingers, flicking fingers, making spiderman pose, \etc.
Some gestures have zero misclassification during the study, such as Snap, Spiderman, Clap, and FiveFingerBend.}

\review{
Meanwhile, false-positive rate was kept low. A few participants tried diverse non-gesture motion and the real-time system was robust to noisy data. On average, participants had 0.6 times of false positive throughout the whole study.}

\review{
Table~\ref{tab:realtime_performance} summarizes the proposed gestures and their recognition performance.
These results indicate the scalability and the robustness of our framework.
}

\subsubsection{Procedure of Gesture Customization}
\review{
During the study, all participants added the RotateOut gesture smoothly.
As for adding their own gestures, most participants succeeded at the first trial when adding their own gesture. This indicates that the system's sanity check does not pose much restriction on users' creativity.}

\review{
A small number of participants' new gesture did not went through due to the similarity check.}
Two participants' first gestures (middle finger pinching and hang loose) were recognized as close to Pinch and Clench, respectively.
P15 attempted to add a slow waving motion as the second gesture but it got recognized as being close to common daily activities.
P18 first added PinkyPinch as the first new gesture, and then tried to add a thumb-tapping on index finger knuckle. It did not go through as it was recognized as being close to PinkyPinch.
\review{
Moreover, some participants deliberately tried to confuse the system. After adding the first gesture, P2 intentionally added another similar gesture but it did not go through.
The accelerometer signals of these new gestures and the wrongly recognized gestures were indeed similar. On the one hand, this indicates the robustness of the system; On the other hand, this reveals the room for improvement of our framework to distinguish closer gestures.}

\review{
As for the model performance check, P1 and P19 got an accuracy below 80\% when adding their gestures. Both of them chose to collect more shots and completed the addition.
}

\renewcommand{\arraystretch}{1.3}
\begin{table*}[t]
\centering
\centering
\resizebox{1\textwidth}{!}{
\begin{tabular}[t]{ccccc|ccccc}
\hline
\hline
\textbf{Gestures}  & \textbf{acc}    & \textbf{F1}    & \textbf{FP Count}  & \textbf{\# of Users} & \textbf{Gestures}  & \textbf{acc}    & \textbf{F1}    & \textbf{FP Count}  & \textbf{\# of Users} \\ \hline
Clench & 0.950 & 0.972 & 0.100 & 20 & DoubleClench & 0.993 & 0.996 & 0.050 & 20\\
Pinch & 0.975 & 0.986 & 0.200 & 20 & DoublePinch & 0.952 & 0.972 & 0.000 & 20\\ \cline{1-10}
RotateOut & 0.949 & 0.973 & 0.050 & 20 & Snap & 0.960 & 0.978 & 0.000 & 5\\
Flick & 0.958 & 0.978 & 0.000 & 3 & Peace & 1.000 & 1.000 & 0.500 & 2\\
FastArmWave & 0.909 & 0.950 & 0.500 & 2 & Spiderman & 1.000 & 1.000 & 0.000 & 1\\
VulcanSalute & 1.000 & 1.000 & 0.000 & 1 & Clap & 1.000 & 1.000 & 0.000 & 1\\
PinchThenDrag & 1.000 & 1.000 & 0.000 & 1 & FiveFingerBend & 1.000 & 1.000 & 0.000 & 1\\
PinkyPinch & 0.900 & 0.947 & 0.000 & 1 & IndexFingerPointing & 0.882 & 0.938 & 0.000 & 1\\
ThumbsUp & 0.857 & 0.923 & 1.000 & 1 & ThumbTwoTaps & 0.833 & 0.909 & 0.000 & 1\\
FiveFingerSequentialBend & 0.812 & 0.897 & 0.000 & 1 & IndexFingerBend & 0.800 & 0.889 & 0.000 & 1\\
Punch & 0.778 & 0.875 & 2.000 & 1 & IndexMiddleFingerBend & 0.750 & 0.857 & 0.000 & 1\\
\hline
\hline
\end{tabular}
}
\vspace{0.1cm}
\caption{Recognition Results in The Real-time System Usability Study. All new gestures are added with three shots. FP count indicates the average count of false positive per person during the study.}
\label{tab:realtime_performance}
\Description{
Results of all gestures were evaluated by participants during the user study. The table shows the gesture name, accuracy, F1 score, false-positive count at the gesture level, as well as the number of participants evaluated it.
}
\end{table*}
\renewcommand{\arraystretch}{1.0}

\subsubsection{Questionnaire Results}
The questionnaire results also suggest positive feedback from participants.
Following the score calculation method~\cite{bangor_empirical_2008}, we obtain the average overall SUS score as 87.2$\pm$8.3 out of 100. This indicates the high overall usability of our system.
SUS has two sub-scale scores (Q4 and Q10 for ``Learnability'' and the rest for ``Usability''). The learnability score was 84.5$\pm$14.3 (out of 100, high) and the usability score was 87.9$\pm$8.6 (out of 100, high). Both sub-scale scores further indicate that our system is easy-to-learn and easy-to-use. The details of each SUS questions can be found in Appendix Figure~\ref{fig:sus}.

\review{Moreover, results of the NASA-TLX questionnaire (on a 7-point Likert scale) indicate that participants had low task load when using the real-time system, which is in line with the SUS outcome. Participants reported low task mental load (1.9$\pm$1.1), physical load (2.7$\pm$1.3), and temporal load (2.3$\pm$1.3). They considered themselves paying low effort (2.0$\pm$0.9). Moreover, participants agreed that they had a good performance during the study (6.2$\pm$1.0) and there was little frustration (1.5$\pm$1.1).
Both the SUS and NASA-TLX questionnaires' results indicate good usability of our system.
}

\subsubsection{Subjective Feedback}
In addition to the questionnaire results, participants provided positive comments about the system.
8 out of the 20 participants explicitly mentioned that they would love to use the system in daily life.
A few participants were impressed by the system: \textit{``It works amazingly well! That's beyond my expectation.''} (P11). Some participants were happy to see how the system can be robust against noisy motion.
P2 was excited when their confusing gesture was not accepted by the system. \textit{``The system could tell that they are too close and it did not let me add it. This is a very good design, [making the system] much more robust!''} (P2).
P1 liked the feedback when the system has sub-optimal performance. \textit{``Know that the system is not perfect is absolutely fine! It only took three samples! I also feel good that the system can inform me about the performance so that I can have a better expectation.''} (P1).
Some participants discussed a few potential use cases of our system. We will have more discussion in Section~\ref{sub:discussion:application}.

Overall, the recognition performance and subject feedback of the user study illustrate the good usability and promising future of our framework.
\section{Discussion}
\label{sec:discussion}
Here, we discuss \review{insights of extending gesture sets}, application scenarios and the potential generalizability of our framework. We also summarize limitations and future work.

\subsection{Potentials and Challenges of Extending Gesture Set}
\review{
Through our evaluation studies, we reveal the potential of our framework to extend the existing four gestures to support more gestures with few shots. In Section~\ref{sec:study_algorithm}, the 12 new gestures are designed to span across different dimensions in hand gesture design space. Meanwhile, in Section~\ref{sec:study_usability}, a number of diverse gestures were proposed by participants without restrictions. Achieving good results in both studies, our framework shows the ability to extend to a wide range of gesture sets.
However, we also foresee some challenges of the framework. Our pilot study indicates that gestures with the same fingers but opposite sequences are hard to classify, \eg thumb sliding from the bottom to the top of the index finger \vs thumb sliding from the top to the bottom of the index finger. This is mainly caused by the similar motion patterns measured from the wrist-worn accelerometer. Increasing the signal sampling rate could be a potential solution to discover pattern differences and distinguish this type of complex but close gestures~\cite{laput_viband_2016}.
}

\subsection{Applications}
\label{sub:discussion:application}
We believe our contributions in this paper can be impactful in many areas.
First, a well-designed customization system can improve gesture memorability~\cite{nacenta_memorability_2013} and interaction efficiency~\cite{ouyang_bootstrapping_2012}. Likewise, our work can be helpful for users who need more than what typically ships in a pre-packaged gesture set. A well-designed customization system can accelerate users' ability to easily and creatively add new gestures.
In our case, we envision our system being particularly helpful for users who have personalized accessibility conditions~\cite{anthony_analyzing_2013} or situational impairments~\cite{sears_when_2003,goel_walktype_2012}. In situations where the original gesture set can be inappropriate or inaccessible, our framework can support the creation of gestures that best cater to users' preferences and abilities; our interactive feedback mechanisms ensure that end-users get to decide what level of robustness and accuracy helps them achieve their device usage goals.

\subsection{Beyond Gestures}
\label{sub:discussion:beyond_gesture}
Our framework has the potential to transfer to other customization tasks beyond gesture recognition.
As depicted in Figure~\ref{fig:overview_framework}, our model architecture and data processing techniques are mostly independent from any specific classification task.
For example, our framework can be applied to other time-series recognition tasks, such as facial expression recognition~\cite{wu_adaptive_2018}, voice command recognition~\cite{hershey_cnn_2017}, and human activity recognition~\cite{gong_metasense_2019}.
As long as the model can be architecturally decomposed into feature extraction and inference components (which is often the case for deep-learning models), the core ideas, interactive feedback mechanisms, and overall contributions in this paper are conceptually and practically compatible.

\subsection{Limitations}
\label{sub:discussion:limitation}
Like any other paper, our work has limitations.
First, our set of 12 new customized gestures is not comprehensive. Although our evaluation is based on all possible combinations of these gestures, our results are still far from being thorough. In the future, we plan to collect data from more gestures and conduct a wider evaluation experiment.
Second, the constraints of our usability study prevented us from investigating the robustness of the system when running for a longer period. It is possible that after a while, users' customized gestures may drift over time. However, we envision multiple techniques to address such variations. For example, when a misclassification is noticed (due to temporal drift), users can provide \textit{in-situ} feedback (via extra gesture samples), helping the system adaptively improve its robustness.
Third, on-device processing and training is beyond the scope of this paper. Currently, our model is trained on an external laptop (with data streamed wirelessly). In an engineering implementation, training and processing can be offloaded to a cloud server, or it can be federated across other devices~\cite{li_federated_2020}. These are areas we plan to investigate in future work.

\section{Conclusion}
\label{sec:conclusion}

In this paper, we present a gesture customization framework that supports end-users to add their own customized gestures with very few samples, without impacting the recognition performance of the existing gesture set.
We first conducted a large-scale user study (N=512) to train an IMU-only deep learning gesture recognition model that can recognize four gestures (Clench, Double-Clench, Pinch, and Double-Pinch) with a cross-user accuracy of 95.7\% and a F1 score of 95.8\% and a false positive rate of 0.6 times per hour.
Then, we proposed a dynamic few-shot learning framework that creates a branch after the first half of the pre-trained model to enable knowledge transfer and introduce minimal influence on the old gestures' recognition outcome. We then used a series of data processing techniques to improve the robustness of the additional prediction model. 
Through an evaluation study (N=20) on a set of 12 new gestures, our framework shows an average accuracy of 55.3\%, 83.1\%, and an F1 score of 66.0\%, 89.2\%, and 92.1\% on using one, three, five shots when adding one new gesture.
When adding two, three, and four gestures, it can achieve an average accuracy of 83.3\%, 77.7\%, and 77.2\% and an average F1 score of 88.8\%, 84.2\%, and 83.4\% with only three shots, while maintaining the low false positive rate and the good accuracy on the existing four gestures.
We further evaluated the usability of the real-time implementation of our framework via a user study (N=20). The results indicate good learnability and usability of our framework.
We envision our work can paves the way for enabling users move beyond pre-existing gestures, freeing them to creatively add new gestures that are tailored to their preferences and ability.

\begin{acks}
We would like to thank all participants for contributing our studies, and all study coordinators for hosting the studies. We also thank Joseph Chen for providing valuable suggestions on ML techniques.
\end{acks}

\balance{}

\bibliographystyle{ACM-Reference-Format}
\bibliography{main}

\newpage

\section*{Appendix}

\renewcommand{\arraystretch}{1.3}
\begin{table*}[b]
\centering
\resizebox{1\textwidth}{!}{
\begin{tabular}[t]{c|cc|cc|cc|cc|cc|cc|c|c}
\hline
\hline
\multirow{3}{*}{$\begin{array}{@{}c@{}}\textbf{\#}\, \textbf{\hbox{of}}\\ \textbf{\hbox{Shots}}\end{array}$} & \multicolumn{4}{c|}{$\begin{array}{@{}c@{}}\textbf{\hbox{New Gestures}}\\\textbf{\hbox{with Prediction Head}}\end{array}$}                & \multicolumn{4}{c|}{$\begin{array}{@{}c@{}}\textbf{\hbox{New Gestures}}\\\textbf{\hbox{with Both Models}}\end{array}$}& \multicolumn{4}{c|}{$\begin{array}{@{}c@{}}\textbf{\hbox{New}}\, \textbf{\&}\, \textbf{\hbox{Existing Gestures}}\\\textbf{\hbox{with Both Models}}\end{array}$} & \multicolumn{2}{c}{\textbf{Non-gestures}} \\\cline{2-15}
 & \multicolumn{2}{c|}{\textbf{Window-level}} & \multicolumn{2}{c|}{\textbf{Gesture-level}} & \multicolumn{2}{c|}{\textbf{Window-level}}    & \multicolumn{2}{c|}{\textbf{Gesture-level}}   & \multicolumn{2}{c|}{\textbf{Window-level}}               & \multicolumn{2}{c|}{\textbf{Gesture-level}}             & \textbf{Window-level}   & \textbf{Gesture-level}   \\ \cline{2-15}
 & \textbf{acc}             & \textbf{F1}             & \textbf{acc}             & \textbf{F1}              & \textbf{acc}              & \textbf{F1}               & \textbf{acc}              & \textbf{F1}               & \textbf{acc}                    & \textbf{F1}                    & \textbf{acc}                   & \textbf{F1}                    & \textbf{FP Rate}        & \textbf{FP Count/Hr}     \\ \hline
1 & 0.908 & 0.570 & 0.523 & 0.608 & 0.889 & 0.434 & 0.505 & 0.592 & 0.880 & 0.546 & 0.680 & 0.709 & 0.001 & 0.040 \\
2 & 0.920 & 0.672 & 0.694 & 0.767 & 0.904 & 0.531 & 0.671 & 0.748 & 0.889 & 0.604 & 0.782 & 0.811 & 0.001 & 0.060 \\
3 & 0.927 & 0.725 & 0.780 & 0.842 & 0.911 & 0.584 & 0.753 & 0.821 & 0.893 & 0.632 & 0.833 & 0.860 & 0.001 & 0.074 \\
4 & 0.930 & 0.747 & 0.824 & 0.875 & 0.914 & 0.613 & 0.798 & 0.855 & 0.895 & 0.645 & 0.861 & 0.884 & 0.001 & 0.102 \\
5 & 0.930 & 0.755 & 0.841 & 0.887 & 0.915 & 0.626 & 0.816 & 0.869 & 0.896 & 0.650 & 0.872 & 0.893 & 0.002 & 0.116 \\
6 & 0.929 & 0.760 & 0.857 & 0.899 & 0.914 & 0.636 & 0.832 & 0.880 & 0.894 & 0.652 & 0.881 & 0.900 & 0.002 & 0.152 \\
7 & 0.931 & 0.764 & 0.858 & 0.900 & 0.916 & 0.636 & 0.832 & 0.881 & 0.896 & 0.654 & 0.882 & 0.901 & 0.002 & 0.145 \\
8 & 0.933 & 0.770 & 0.864 & 0.904 & 0.917 & 0.641 & 0.838 & 0.886 & 0.897 & 0.657 & 0.885 & 0.904 & 0.002 & 0.128 \\
9 & 0.931 & 0.770 & 0.870 & 0.909 & 0.916 & 0.644 & 0.843 & 0.890 & 0.896 & 0.657 & 0.888 & 0.906 & 0.002 & 0.136 \\
10 & 0.932 & 0.778 & 0.885 & 0.919 & 0.917 & 0.657 & 0.860 & 0.901 & 0.896 & 0.662 & 0.899 & 0.915 & 0.002 & 0.162 \\
\hline
\hline
\multirow{3}{*}{$\begin{array}{@{}c@{}}\textbf{\#}\, \textbf{\hbox{of}}\\ \textbf{\hbox{Gestures}}\end{array}$} & \multicolumn{4}{c|}{$\begin{array}{@{}c@{}}\textbf{\hbox{New Gestures}}\\\textbf{\hbox{with Prediction Head}}\end{array}$}                & \multicolumn{4}{c|}{$\begin{array}{@{}c@{}}\textbf{\hbox{New Gestures}}\\\textbf{\hbox{with Both Models}}\end{array}$}& \multicolumn{4}{c|}{$\begin{array}{@{}c@{}}\textbf{\hbox{New}}\, \textbf{\&}\, \textbf{\hbox{Existing Gestures}}\\\textbf{\hbox{with Both Models}}\end{array}$} & \multicolumn{2}{c}{\textbf{Non-gestures}} \\\cline{2-15}
 & \multicolumn{2}{c|}{\textbf{Window-level}} & \multicolumn{2}{c|}{\textbf{Gesture-level}} & \multicolumn{2}{c|}{\textbf{Window-level}}    & \multicolumn{2}{c|}{\textbf{Gesture-level}}   & \multicolumn{2}{c|}{\textbf{Window-level}}               & \multicolumn{2}{c|}{\textbf{Gesture-level}}             & \textbf{Window-level}   & \textbf{Gesture-level}   \\ \cline{2-15}
 & \textbf{acc}             & \textbf{F1}             & \textbf{acc}             & \textbf{F1}              & \textbf{acc}              & \textbf{F1}               & \textbf{acc}              & \textbf{F1}               & \textbf{acc}                    & \textbf{F1}                    & \textbf{acc}                   & \textbf{F1}                    & \textbf{FP Rate}        & \textbf{FP Count/Hr}     \\ \hline
1 & 0.936 & 0.834 & 0.828 & 0.883 & 0.921 & 0.744 & 0.805 & 0.865 & 0.885 & 0.651 & 0.918 & 0.930 & 0.000 & 0.025 \\
2 & 0.934 & 0.785 & 0.824 & 0.875 & 0.917 & 0.664 & 0.800 & 0.857 & 0.891 & 0.648 & 0.886 & 0.904 & 0.001 & 0.062 \\
3 & 0.928 & 0.738 & 0.795 & 0.849 & 0.912 & 0.604 & 0.770 & 0.831 & 0.893 & 0.637 & 0.850 & 0.872 & 0.001 & 0.095 \\
4 & 0.926 & 0.718 & 0.798 & 0.848 & 0.910 & 0.586 & 0.773 & 0.829 & 0.894 & 0.633 & 0.838 & 0.861 & 0.002 & 0.127 \\
\hline
\hline
\end{tabular}
}
\vspace{0.1cm}
\caption{Results Summary of Prediction Heads with Different Numbers of Shots (top) and Gestures (bottom). FP stands for false positive. The top table shows the average results over 1 to 10 shots. The bottom table shows the average results over 1 to 4 gestures.}
\label{tab:fewshots_results}
\Description{
Two tables about the evaluation results. The first table summarizes the results when using different numbers of shots. The second table summarizes the results when using different numbers of gestures. The two tables have a similar structure. Each has four column groups: new gestures with prediction head, new gestures with prediction head and pre-trained models, new and existing gestures with both models, and non-gestures.
For each column group, it shows window-level and gesture-level results. Within each level, it shows accuracy and F1 score.
}
\end{table*}
\renewcommand{\arraystretch}{1.0}

\begin{figure*}[b]
    \centering
    \includegraphics[width=1\textwidth]{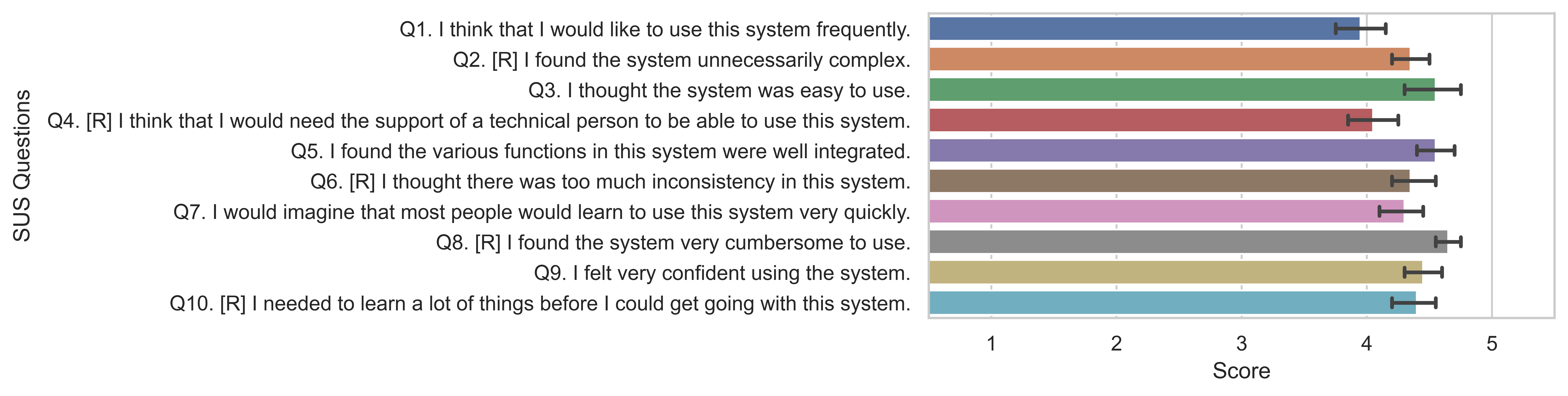}
    \caption{Barplot of the 10 Questions in SUS Questionnaire. Note that Q2,4,6,8,10 are marked as [R] and their scores are reversed for better visualization. Error bar indicates standard error.}
    \label{fig:sus}
    \Description{
A horizontal bar plot of SUS results. The X-axis is the scores from 1 to 5, while Y-axis shows the 10 SUS questions:
 Q1. I think that I would like to use this system frequently.
 Q2. I found the system unnecessarily complex.
 Q3. I thought the system was easy to use.
 Q4. I think that I would need the support of a technical person to be able to use this system.
 Q5. I found the various functions in this system were well integrated.
 Q6. I thought there was too much inconsistency in this system.
 Q7. I would imagine that most people would learn to use this system very quickly.
 Q8. I found the system very cumbersome to use.
 Q9. I felt very confident using the system.
Q10. I needed to learn a lot of things before I could get going with this system.
The scores of Q2, Q4, Q6, Q8, and Q10 are reversed in the plot to enable consistent visualization.
}
\end{figure*}

\end{document}